\documentclass[11pt,a4paper]{article}

\usepackage[english]{babel} 
\usepackage{dsfont} 
\usepackage{amsmath,amsfonts,amsthm}
\usepackage{a4wide}
\usepackage{fancyhdr}
\usepackage{latexsym}
\usepackage{amsmath,amssymb,amsfonts}%
\usepackage{paralist}
\usepackage{lscape}
\usepackage{url}
\usepackage{xfrac}

\usepackage{graphicx}
\usepackage{pifont}
\newcommand{\cmark}{\ding{51}}
\newcommand{\xmark}{\ding{55}}

\usepackage{amsthm}%
\usepackage{mathrsfs}%
\usepackage{bm}
\usepackage[title]{appendix}%
\usepackage{xcolor}%
\usepackage{textcomp}%
\usepackage{manyfoot}%
\usepackage{booktabs}%
\usepackage{algorithm}%
\usepackage{algorithmicx}%
\usepackage{algpseudocode}%

\usepackage[numbers,sort&compress]{natbib}
\usepackage{threeparttable}

\usepackage{float}
\usepackage{verbatim}
\usepackage{longtable}
\usepackage{tabularx}       
\usepackage{multicol}
\usepackage{caption}
\usepackage{floatflt}
\usepackage{afterpage}
\usepackage{moreverb}     
\usepackage{bbm}
\usepackage{booktabs}
\usepackage[colorlinks = true, linkcolor = blue, citecolor = blue, urlcolor = blue, breaklinks = true]{hyperref}
\usepackage{comment}

\usepackage{stackengine} 
\newcommand\ubar[1]{\stackunder[1.2pt]{$#1$}{\rule{.8ex}{.075ex}}}

\usepackage{float} 

\usepackage{multirow}
\usepackage[margin=1in]{geometry}
\usepackage{calc}

\allowdisplaybreaks 

\AtBeginDocument{}  

\newcounter{saveeqn}

 \theoremstyle{definition}
\newtheorem{assumption}{Assumption} 

\newtheorem{theorem}{Theorem}
\newtheorem{lemma}{Lemma} 
\newtheorem{corollary}{Corollary}
\newtheorem{proposition}{Proposition}
\newtheorem{definition}{Definition}
\newtheorem{remark}{Remark}
\newtheorem{example}{Example}

{
\par\vspace{\baselineskip}\noindent
\textbf{Example #1 -- continued.}
}%
{%
\ignorespacesafterend
}

\newcommand{\Var}{\text{Var}}
\newcommand{\Cov}{\text{Cov}}
\newcommand{\argmin}{\text{argmin}}
\newcommand{\argmax}{\text{argmax}}
\renewcommand{\a}{\alpha}
\newcommand{\g}{\gamma}
\renewcommand{\t}{\Theta}
\newcommand{\T}{\mathcal{T}}
\newcommand{\Tc}{{\T^c}}
\newcommand{\R}{\mathbb{R}}
\newcommand{\E}{\mathbb{E}}
\renewcommand{\P}{\mathbb{P}}
\newcommand{\essinf}{\text{ess\,inf}}
\newcommand{\esssup}{\text{ess\,sup}}
\renewcommand{\bm}{\boldsymbol}
\newcommand{\Ind}{\mathds{1}}
\renewcommand{\L}{\mathcal{L}}
\newcommand{\U}{\mathcal{U}}
\newcommand{\V}{\mathcal{V}}
\newcommand{\dG}{\text{d}G}
\newcommand{\dF}{\text{d}F}

\usepackage{enumitem}

\begin{document}


\begin{center}
{\Large {\bf Optimal basis risk weighting in expectile-based parametric insurance}}\\
\vspace*{0.8cm}
{\large Markus Johannes Maier\textsuperscript{1*} and Matthias Scherer\textsuperscript{1}}\\
\vspace*{0.5cm}
\textsuperscript{1}Chair of Mathematical Finance, Technical University of Munich (TUM), Parkring 11, D-85748 Garching-Hochbrück, Germany\\
\vspace*{0.5cm}
\textsuperscript{*}Corresponding author: \href{mailto:markus.j.maier@tum.de}{markus.j.maier@tum.de}\\
Contributing author: \href{mailto:scherer@tum.de}{scherer@tum.de} 
\end{center}
\vspace*{0.5cm}

\begin{abstract}\noindent
Parametric insurance contracts translate index measurements to compensation for policyholders' losses using  predefined payment schemes. These need to be designed carefully to keep basis risk, i.e. the disparity between payouts and true damages, small. Previous research has motivated the use of conditional expectiles as payment schemes, whose compensation is impacted by the policyholder's potentially unknown attitude towards basis risk. To alleviate this model uncertainty and to investigate the impact of (hidden) influencing factors, we characterize existence and uniqueness of the optimal basis risk weighting in a utility-maximization framework through a set of boundary conditions. In the absence of an optimal solution, we provide comparisons to the utility of no insurance and full indemnity coverage. We establish a link between location-scale distributions and separability of conditional expectiles' derivatives, thus improving the understanding of these statistical functionals. A simulation study on parametric hurricane insurance visualizes our results, investigates the influence of premium loading and risk aversion on the optimal weighting, and comments on the challenge of (spatial) loss dependence.

\vspace{0.4cm}
\noindent
\textbf{Keywords:} Basis risk, Expectiles, Parametric insurance, Hurricane risk.
\end{abstract}
 

\section{Introduction}\label{Introduction}
Parametric insurance is an alternative risk transfer solution that is expected to experience a sizable market growth in the coming decade \cite{GMI2025}. First introduced in the agricultural sector as a protection against crop loss, domains of application have widened to include (among others) natural catastrophes, pandemics, construction delays, and business interruption \cite{MunichReWeb, SwissRe2024}. Irrespective of the risk to be covered, parametric insurance's primary characteristic is that compensation is based on publicly observable information (called parameter / index\footnote{Therefore, parametric insurance may also be called ``index insurance'', with the latter often referring to yield loss protection in earlier literature \cite{Lin2020}. While some sources make a more detailed differentiation between both terms (see, e.g., \cite[Comment 18]{FSI2024}), we will use them synonymously.}) and payment schemes instead of the policyholder's true loss \cite{Louaas2024}. Examples for indices include average rainfall in agriculture insurance \cite{Lin2020} and maximum wind speed in hurricane covers \cite{SwissRe2020}. With regards to payment schemes, one distinguishes between pure parametric insurance, where triggered payments are constant, and parametric index insurance, whose compensation depends on the concrete index measurement\footnote{A mathematical description of the admissible payment schemes for both types of parametric insurance is given in Proposition~\ref{PropBROpt}.}~\cite{Lin2020}.

Compared to indemnity insurance, parametric solutions offer beneficial reductions in premium, time until payout, and complexity, since no extensive loss assessment is required after an incident. However, this simplified approach to compensation may result in differences between the true damage and the payout. Such deviations are referred to as \emph{basis risk} and need to be kept low, as they present a notable threat to the viability of parametric insurance \cite{Lin2020}. To that end, many researchers have investigated optimal index insurance design, either in terms of maximizing the policyholder's expected utility (see, e.g., \cite{Mahul1999, Mahul2001,Mahul2003, Bourgeon2003, Zhang2019, Louaas2024}) or by minimizing a chosen basis risk metric (see, e.g., \cite{Brown2007, Steinmann2023, Stigler2023}). Following the latter approach, our prior research in \cite{Maier2025} investigated the nature of basis risk-optimal payment schemes. In an asymmetrically weighted mean square error framework, these were shown to correspond to conditional expectiles (see~\cite{Bellini2018}) of the policyholder’s true loss given a compensation-triggering incident. The expectile's level and therefore the overall payment depends on the relative importance $\alpha$ of negative basis risk (undercompensation) compared to positive basis risk (overcompensation).

While able to provide important insights into contract design and to link parametric insurance to the heavily researched domain of expectiles (see, e.g., \cite{Bellini2015, Delbaen2016, Ziegel2016,Bellini2014, Bellini2018, Girard2020, Daouia2018}), the model in \cite{Maier2025} leaves room for improvement. Specifically, it assumes $\a$ to be known, which introduces some inherent uncertainty. Moreover, while basis risk plays a pivotal role for the viability of parametric insurance contracts, it is clear that a policyholder's attitude towards it may be shaped by factors like risk aversion, personal wealth, and premium cost\footnote{For example, policyholders might accept higher (negative) basis risk if premiums are cheaper.}, which have not been (explicitly) considered in the underlying framework. 

To improve the model's reliability and to investigate hidden influencing factors, we consider the choice of $\a$ in the context of utility-maximization. More precisely, we investigate whether a policyholder’s expected utility with expectile-based payment schemes as derived in \cite{Maier2025} admits a (unique) maximum w.r.t. the basis risk weighting $\a$. Under different premium principles, we are able to characterize existence and uniqueness of the optimal $\a$ through two boundary conditions and can (numerically) derive its value. This major step towards the implementation of expectile-based payment schemes presents our contribution to the growing literature on the optimal design of parametric contracts (see, e.g., \cite{Zhang2019, Stigler2023, Louaas2024}). Our results apply to both pure parametric and parametric index insurance, with the latter requiring a separability assumption on the expectiles' derivatives that turns out to be structurally related to location-scale distribution families. If no optimal weighting exists, parametric contracts’ expected utility can be compared to no insurance resp. full coverage indemnity insurance. 

To visualize our results, we conduct a simulation study on parametric cat-in-a-circle hurricane insurance based on maximum windspeed, inspired by a common area of application for index insurance in practice, see~\cite{SwissRe2020}. Using synthetic hurricane tracks from the STORM\footnote{\textbf{S}ynthetic \textbf{t}ropical cycl\textbf{o}ne gene\textbf{r}ation \textbf{m}odel.} data set \cite{BloemendaalData}, we provide guidance on numerical derivation of the optimal basis risk weighting and investigate the influence of premium loading and risk aversion. We examine piecewise linear payment schemes (commonly employed in real-world contracts, see~\cite{Steinmann2023}) in addition to the theoretically optimal conditional expectiles and briefly touch upon the topic of spatial dependence between hurricane losses.

The rest of the paper is structured as follows. In Section~\ref{Preliminaries}, we summarize important mathematical properties of conditional expectiles as well as the basis-risk optimal payment schemes of \cite{Maier2025}. Section~\ref{TheoOpt} introduces our main results by investigating existence, uniqueness, and numerical derivation of the optimal basis risk weighting in pure parametric (Section~\ref{PP}) and parametric index insurance (Section \ref{PI}) and gives insight into the insurance decision in the absence of an optimal solution. Section~\ref{SimStudy} presents the framework for and the results of our simulation study on parametric hurricane insurance. Our findings and opportunities for future research are summarized in Section~\ref{Conclusion}.

\section{Mathematical preliminaries and notation}\label{Preliminaries}
In the following, we briefly introduce the notation and important results for our derivations in Section \ref{TheoOpt}. All random variables (rv) are defined on a suitable probability space $(\Omega,\mathcal{F},\P)$. We denote by $\mathcal{L}^p(\mathcal{G}) := \mathcal{L}^p(\Omega,\mathcal{G},\P)$, $p \in [1,\infty]$, $\mathcal{G} \subset \mathcal{F}$, the space of $\mathcal{G}$-measurable rvs with $ ||X||_p < \infty$. For convenience, we write $\mathcal{L}^p := \mathcal{L}^p(\mathcal{F})$ and $\mathcal{L}^p(\t) := \mathcal{L}^p\bigl(\sigma(\t)\bigr)$ for some rv $\t$.

Basis risk-optimal payment schemes in the sense of \cite{Maier2025}, which we will formalize in Definition~\ref{DefBROpt} and Proposition~\ref{PropBROpt}, are closely related to conditional expectiles, see~\cite{Bellini2018}. These statistical functionals extend the well-known concept of expectiles, introduced by \cite{Newey1987}, from constants to $\mathcal{G}$-measurable random variables and can be defined as follows.\footnote{Definition~\ref{DefCondExp} can be extended to $\mathcal{L}^1$ rvs. The interested reader may refer to \cite[Definition 1]{Bellini2018}.}

\begin{definition}[Conditional expectiles \cite{Bellini2018}]\label{DefCondExp}
For any $\gamma \in (0,1)$ and $\mathcal{G} \subset \mathcal{F}$, the \emph{conditional expectile} $e_\gamma(X|\mathcal{G})$ of a rv $X \in \mathcal{L}^2$ is $\mathbb{P}$-a.s.\ uniquely given by 
\begin{equation*}\label{EqCondExp}
e_\gamma(X|\mathcal{G}) := \underset{Y \in \mathcal{L}^2(\mathcal{G})}{\mathrm{argmin}} \; \mathbb{E}\left[ \gamma (X-Y)_+^2 + (1-\gamma) (X-Y)_-^2 \right],
\end{equation*}
where  $(x)_+ := \max\{x,0\}$ and $(x)_- := \max\{-x,0\}$. 
\end{definition}
\noindent For generated $\sigma$-algebras, we use the following abbreviations for ease of notation:
\begin{itemize}
\item $e_\gamma\bigl(X\big| \sigma(A)\bigr) =: e_\gamma\bigl(X\big| A \bigr)$ for a set $A \in \mathcal{F}$,
\item $e_\gamma\bigl(X\big| \sigma(\t) \bigr)(\omega) =: e_\gamma\bigl(X\big| \t \bigr)(\omega)  =: e_\gamma\bigl(X\big| \theta \bigr)$ for a rv $\t$ and $\omega \in \Omega$ with $\t(\omega) = \theta$.\footnote{Recall that $e_\gamma\bigl(X\big| \sigma(\t) \bigr)$ is $\sigma(\t)$-measurable.}
\end{itemize}

Conditional expectiles have appealing properties, see Lemma~\ref{PropCondExp}. Especially, Property~\ref{ExpToCond} enables us to transfer results from the domain of (unconditional) expectiles $e_\g(X) := e_\g(X|\{\emptyset, \Omega\})$ to the conditional setting. To avoid meritlessly complicating formulas, Theorem \ref{PropExp} presents these properties in terms of unconditional expectiles. 

\begin{lemma}[Properties of $e_\gamma(\circ|\mathcal{G})$ \cite{Bellini2018}] \label{PropCondExp}
For any $\gamma \in (0,1)$ and $X \in \mathcal{L}^2$, it holds $\mathbb{P}$-a.s.:  
\begin{enumerate}
\item $e_\gamma( \Lambda X+H|\mathcal{G}) =\Lambda e_\gamma(X|\mathcal{G}) + H$ for $H \in \mathcal{L}^2(\mathcal{G})$ and non-negative \mbox{$\Lambda \in \mathcal{L}^\infty(\mathcal{G})$}. \label{CondCI}
\item $e_\gamma(X|\mathcal{G})(\omega) = e_\gamma\bigl(F_\mathcal{G}(\circ,\omega)\bigr)$, $\omega \in \Omega$, where $F_\mathcal{G}(\circ,\omega)$ is a regular conditional distribution of $X$ on $\mathcal{G}$. \label{ExpToCond}
\end{enumerate}
\end{lemma}
\begin{theorem}[Properties of $e_\g(\circ)$ \cite{Newey1987, Jones1994, Bellini2016}] \label{PropExp} 
For non-deterministic $X \in \L^2$, $X \sim F$, it holds:
\begin{enumerate}
\item $(0,1) \ni \g \mapsto e_\g(X)$ is strictly increasing and continuous. \label{MonoExp}
\item $\lim_{\g \searrow 0} e_\g(X) = \essinf(X)$, $\lim_{\g \nearrow 1} e_\g(X) = \esssup(X)$. \label{LimitExp}
\item If $F$ is continuously differentiable, then $\g \mapsto e_\g(X)$ is continuously differentiable and \label{DiffReq}
\begin{equation}
F\bigl(e_\g(X) \bigr) = - \frac{e_\g(X) - \E[X] + \g(1-2\g) e'_\g(X)}{(1 - 2\g)^2 e'_\g(X)}, \label{ExpDerivative} 
\end{equation}
for any $\g \neq 1/2$ as well as in the limit for $\g = 1/2$, or equivalently
\begin{equation}
e'_\g(X) = \frac{\E\bigl[ |X - e_\g(X)| \bigr]}{(1 - \g)F\bigl(e_\g(X)\bigr) + \g \bar F\bigl(e_\g(X)\bigr)}, \label{AltExpDerivative} 
\end{equation}
for all $\g \in (0,1)$. Here, $\bar F(x) = 1 - F(x)$.
\item For any $\g \in (0,1)$, $e_\g(X)$ is the unique $y$ solving
\begin{equation}  \label{ExpSolve}
\gamma = \frac{L(y) - yF(y)}{2\bigl(L(y) - yF(y)\bigr) + y - \E[X]},
\end{equation}
with $L(y) := \int_{-\infty}^y x\, \dF(x)$.
\end{enumerate}
\end{theorem}

Next, let us summarize the approach towards optimal basis risk from \cite{Maier2025}. For a given policyholder, we investigate a parametric coverage for a loss $S \in \mathcal{L}^2$ with cdf $F$. A deterministic payment scheme provides compensation based on the \emph{index} $\t \sim G$ and the \emph{trigger area} $\mathcal{T} \subset \t(\Omega)$, which fulfills \mbox{$\P(\t \in \T) \in (0,1)$} and contains all values of $\t$ that result in a payout. $\t$ and $\T$ are assumed to be outside the policyholder's control.\footnote{While parametric contracts can easily be customized, the choice of index and trigger area depends on considerations of scalability as well as on availability and granularity of public information with strong correlation to the severity of the loss $S$.} Additionally, payment schemes are required to result in square-integrable payments $Y \in \mathcal{L}^2(\t)$ to match the tail-behaviour of $S$. 

For the above contract, basis risk is measured as the asymmetrically weighted mean squared difference between losses and payments and depends on the basis risk weighting $\a \in (0,1)$. $\a$ describes the importance of negative basis risk $(S - Y)_-$ relative to positive basis risk $(S - Y)_+$ and is assumed as given in \cite{Maier2025}. As a result, basis risk-optimal payment schemes in the sense of Definition \ref{DefBROpt} are a.s. equal to conditional expectiles of the policyholder's true loss $S$ given the occurrence of a triggering incident, see Proposition \ref{PropBROpt}. Note the slight abuse of language in which we use the term ``payment scheme'' to refer to the payouts $Y$. 

\begin{definition}[Basis risk-optimal payment scheme \cite{Maier2025}]\label{DefBROpt}
For a given set of admissible payment schemes $\bm{Y} \subset \L^2(\t)$ and a fixed basis risk weighting $\a \in (0,1)$, we call a minimizer $Y^*$ of
\begin{equation*}
\min_{Y \in \bm{Y}} \, \E\left[\bigl( \a (S - Y )_+\bigr)^2 + \bigl( \left(1-\a\right) (S - Y)_-\bigr)^2 \right]
\end{equation*}
a \emph{basis risk-optimal payment scheme}.
\end{definition}

\begin{proposition}[Expectiles as basis risk-optimal payment schemes \cite{Maier2025}]\label{PropBROpt}
Let $\T \subset \t(\Omega)$ with $\P(\t \in \T) > 0$ and $\a \in (0,1)$. With $\g(\a) := \frac{\a^2}{(1-\a)^2 + \a^2}\in(0,1)$, the basis risk-optimal payment scheme in the sense of Definition \ref{DefBROpt} under:
\begin{itemize}
\item \emph{Pure parametric insurance} with admissible payment schemes $\bm{Y} = \left\{y\, \Ind_{\{\t\in \T\}} \middle|\, y \in \R_{>0}\right\}$ is a.s. equal to
\begin{equation}\label{PS_PP}
Y^*_{\g(\a)} = e_{\g(\a)}(S|\t \in \T) \, \Ind_{\{\t \in \T\}}.
\end{equation} 
\item \emph{Parametric index insurance} with admissible payment schemes $\bm{Y} = \{Y \, \Ind_{\{\t\in \T\}} |\, Y \in \L^2(\t), Y \ge 0 \}$ is a.s. equal to
\begin{equation}\label{PS_PI}
Y^*_{\g(\a)} = e_{\g(\a)}(S|\t) \, \Ind_{\{\t \in \T\}}.
\end{equation} 
\end{itemize}
\end{proposition}

As pointed out in \cite{Maier2025}, the assumption of $\a$ being known introduces some undesirable (inherent) uncertainty into the model and should be relaxed. In this case, however, identifying an appropriate value of $\a$ for a given policyholder is of obvious importance for implementing the payment schemes (\ref{PS_PP}) and (\ref{PS_PI}). This task represents our core research question in the next section.

\section{Optimal basis risk weighting}\label{TheoOpt}

We assume $\a \in (0,1)$ to be (a priori) unknown to the insurer. Naturally, as $\a$ is the sole quantity \emph{explicitly} describing the policyholder's perception of basis risk, it should encapsulate all influencing factors not (directly) considered in (\ref{DefBROpt}), like risk aversion and premium. Thus, a rational way towards identifying $\a$ is to make these hidden effects tangible by considering the policyholder's ultimate goal of utility-maximization. More precisely, our aim is to characterize existence and uniqueness of the (utility-)optimal basis risk weighting
\begin{equation}\label{EqOpt}
\a^{*} := \underset{\a \in (0,1)}{\argmax} \; \underbrace{\E\left[u(w_0 - S + Y^*_{\g(\a)} - \pi_{\g(\a)}) \right]}_{=: \, \U\bigl(\g(\a)\bigr)},
\end{equation}
where $u: \R \to \R$ is a twice differentiable, concave von Neumann--Morgenstern utility function and $\pi_{\g(\a)}$ is calculated using the expected value ($\text{E}$), standard deviation ($\text{SD}$), or variance ($\text{V}$) premium principle with loading $\rho > 0$:
\begin{align}
\pi^\text{E}_{\g(\a)} &= (1+\rho)\E\left[Y^*_{\g(\a)}\right]. \label{EV_Premium} \\
\pi^{\text{SD}}_{\g(\a)} &= \E\left[Y^*_{\g(\a)}\right] + \rho \, \sqrt{ \Var\left( Y^*_{\g(\a)} \right) }. \label{SD_Premium} \\
\pi^\text{V}_{\g(\a)} &= \E\left[Y^*_{\g(\a)}\right] + \rho \, \Var\left( Y^*_{\g(\a)} \right). \label{Var_Premium}
\end{align}

In the interest of readability, Table~\ref{VariableNotation} summarizes the notation. Additionally, we omit the dependence of $\g$ on $\a$ unless we want to specifically stress this relationship, using the strict increasingness of $\g(\a)$ to recover $\a$ from given $\g \in (0,1)$ as
\begin{equation} \label{alphagamma}
\a = \frac{1}{2}\mathds{1}_{\{\g = \frac{1}{2}\}} + \frac{\g - \sqrt{\g - \g^2}}{2\g - 1}\mathds{1}_{\{\g \neq \frac{1}{2}\}},
\end{equation}
and write $\g^* := \g(\a^*)$. Additionally, we will denote the (partial) derivative of any quantity $Q(\g,\dots)$ w.r.t. $\g$ by $Q'(\g,\dots)$. For completeness' sake, we note that (\ref{EqOpt}) implicitly assumes that at most one incident takes place during the coverage period of the parametric contract. We believe this to be a realistic assumption, especially in the common context of yearly\footnote{While parametric contracts may have a multi-year runtime \cite{SwissRe2024}, this still fits the above setting under the small assumption that the distributional characteristics of ($\t,S$) and the chosen payment scheme remain the same over the period of interest.} yield loss in agriculture or when insuring a building against destruction by natural catastrophes. As introducing the possibility of multiple losses occurring during the coverage period is decidedly non-trivial\footnote{For example, note that expectiles are either sub- or super-additive, depending on their level $\g$, see, e.g., \mbox{\cite[Theorem 1]{Bellini2016}}.}, we will not comment on this issue further.

\begin{table}[t]
\begin{center}
\begin{tabular}{c|l}
\toprule
\textbf{Variable} & \textbf{Meaning}  \\
\midrule
$S$ & True loss \\
$\t$ & Index\\
$F_\theta$ & Conditional cdf of $S$ given $\t = \theta$\\
$G$ & cdf of $\t$ \\ 
$\T$ & Trigger area, i.e. a payment is due if $\t \in \T$\\
$\alpha$ & Importance of negative basis risk relative to positive basis risk \\
$\gamma$ & Level of the optimal conditional expectile governing payment for given $\alpha$ \\
$Y^*_{\gamma}$ & Basis risk-optimal payment scheme for given $\alpha$ \\
$\pi_{\gamma}$ & Premium for parametric contract with payment scheme $Y^*_{\gamma}$ \\
\bottomrule
\end{tabular}
\caption{Important variables introduced in Sections~\ref{Preliminaries} and~\ref{TheoOpt} and their meaning.}
\label{VariableNotation}
\end{center}
\end{table}

To derive the maximizer in (\ref{EqOpt}), it will prove beneficial to consider the behaviour of $\U(\g)$ in the two underlying scenario sets $\T$ (compensation) and $\T^c := \t(\Omega) \big\backslash \T$ (no compensation). Denoting the conditional distribution of $S$ given $\t = \theta$ by $F_{\theta}$, we set
\begin{align*}
\U(\g) =& \,  \int \int u\bigl(w_0 - s + Y^*_{\g} -\pi_{\g} \bigr) \, \dF_{\theta}(s) \, \dG(\theta) = \U_1(\g) + \U_2(\g), \quad \text{where}\\
\U_1(\g) :=&\, \int_\T \int u\bigl(w_0 - s + Y^*_{\g} -\pi_{\g} \bigr) \, \dF_{\theta}(s) \, \dG(\theta), \\
\U_2(\g) :=&\, \int_{\T^c} \int u\bigl(w_0 - s -\pi_{\g} \bigr) \, \dF_{\theta}(s) \, \dG(\theta).
\end{align*}
As $\g^*$ must fulfill the classical condition of a critical point
\begin{align*}
\U'(\g^*) = 0 \quad\Leftrightarrow \quad \U'_1(\g^*) = - \U'_2(\g^*),
\end{align*} 
our aim is to derive monotonicity properties of 
\begin{align*}
\U'_1(\g) &= \int_\T \bigl[(Y^*_{\g})' - \pi'_\g\bigr] \int u'\bigl(w_0 - s + Y^*_{\g} -\pi_{\g} \bigr) \, \dF_{\theta}(s) \, \dG(\theta), \quad \text{and}\\
\U'_2(\g) &= - \int_\Tc \pi'_\g  \int u'\bigl(w_0 - s -\pi_\g \bigr) \, \dF_{\theta}(s) \, \dG(\theta)
\end{align*}
to characterize the existence and uniqueness of $\g^*$. The above requires interchangeability of integration and differentiation. In the setting of Sections \Ref{PP} and \ref{PI}, this is, e.g., guaranteed if marginal utilities are bounded from above on the support of \mbox{$w_0 - S + Y^*_{\g} - \pi_{\g}$}. The interested reader may note the connection to the class of utility functions used in an alternative definition of almost first order stochastic dominance, see \cite{DeVecchi2026}. Differentiability of $Y^*_{\g}$ is guaranteed in the context of Proposition~\ref{PropBROpt} as long as the underlying (regular) conditional distributions are non-atomic (see Theorem~\ref{PropExp}, Property~\ref{DiffReq}), which seems quite realistic.

Following the structure outlined in Proposition \ref{PropBROpt}, we first investigate the case of pure parametric insurance in Section \ref{PP}. Here, uniqueness is always ensured in case of existence, which can be characterized by two boundary conditions (see Theorem~\ref{ExistencePP}). In the context of parametric index insurance, treated in Section \ref{PI}, similar results (see Theorem~\ref{ExistencePI}) can be derived under a distributional assumption on the policyholder's conditional losses that is closely tied to location-scale families of distributions.

\subsection{Pure parametric insurance}\label{PP}

First, let us recall that the optimal payment scheme under pure parametric insurance is given by (\ref{PS_PP}). Now, we can easily see 
\begin{align*} 
\pi^\text{E}_{\g} &= \underbrace{(1+\rho)\P(\t \in \T)}_{=: \, c^\text{E}}e_\g(S| \t \in \T) ,\\
\pi^{\text{SD}}_{\g} &= \underbrace{ \Bigl(\P(\t \in \T) + \rho \, \sqrt{\P(\t \in \T) \, \bigl( 1 - \P(\t \in \T) \bigr)}}_{=: \, c^{\text{SD}}} \Bigr) e_\g(S| \t \in \T),\\
\pi^\text{V}_{\g} &=e_\g(S| \t \in \T) \, \P(\t \in \T) + e^2_\g(S| \t \in \T) \rho \, \P(\t \in \T) \bigl( 1 - \P(\t \in \T) \bigr).
\end{align*}
Especially, note that we can w.l.o.g. assume $c^\circ< 1$ for $\circ \in \{\text{E},\text{SD}\}$ in any realistic setting, as premiums would otherwise a.s. dominate payments. Starting with the expected value and standard deviation principles, we use the fact that $Y^*_{\g}$ is constant on $\T$ to write
\begin{align*}
\U'_1(\g) &= \P(\t \in \T)e'_\g(S|\t \in \T) \E\left[u'\left(w_0 - S +  \bigl(1 - c^{\circ}\bigr) e_\g(S|\t \in \T) \right)| \t \in \T \right]  \bigl(1 - c^\circ\bigr), \\
\U'_2(\g) &=-\P(\t \notin \T) c^\circ e'_\g(S|\t \in \T) \E\left[u'\bigl(w_0 - S - c^\circ e_\g(S|\t \in \T)  \bigr)| \t \notin \T \right],
\end{align*}
for $\circ \in \{\text{E},\text{SD}\}$, which translates to
\begin{align} \label{CritPPESD}
\U'(\g^*) = 0 &\Leftrightarrow \underbrace{\P(\t \in \T) \,\bigl(1 - c^\circ \bigr)\E\left[u'\left(w_0 - S +  \bigl(1 - c^\circ\bigr) e_{\g^*}(S|\t \in \T) \right)| \t \in \T \right]}_{=: \, \mathcal{V}_1(\g^*)} \nonumber \\
&\qquad = \underbrace{\P(\t \notin \T) \, c^\circ \E\left[u'\bigl(w_0 - S - c^\circ e_{\g^*}(S|\t \in \T)\bigr) | \t \notin \T \right]}_{=: \, \mathcal{V}_2(\g^*)}. 
\end{align}
Monotonicity of expectiles (see Therorem \ref{PropExp}, Property \ref{MonoExp}) and concavity of $u$ ensure that $\mathcal{V}_1$ (resp.\ $\mathcal{V}_2$) are strictly decreasing (resp.\ increasing) in $\g$. 

Likewise, if the variance principle is used, $\g^*$ must fulfill
\begin{align} \label{CritPPVar}
&\underbrace{\P(\t \in \T) \,\bigl(1 -r(\g^*) \bigr)\E\left[u'\left(w_0 - S +  \bigl(1 - R(\g^*) \bigr) e_{\g^*}(S|\t \in \T) \right)| \t \in \T \right]}_{=: \, \mathcal{V}_1(\g^*)}  \\
&\hspace{3cm} = \underbrace{\P(\t \notin \T) \, r(\g^*) \E\left[u'\bigl(w_0 - S - R(\g^*) e_{\g^*}(S|\t \in \T)\bigr) | \t \notin \T \right]}_{=: \, \mathcal{V}_2(\g^*)}, \nonumber\\
& R(\g) := \P(\t \in \T) \bigl(1 + \rho \P(\t \notin \T) e_\g(S|\t \in \T)  \bigr), \nonumber \\
& r(\g) := \P(\t \in \T) \bigl(1 + 2\rho \P(\t \notin \T) e_\g(S|\t \in \T)  \bigr). \nonumber
\end{align}
While less obvious to recognize in this case, $\mathcal{V}_1$ (resp.\ $\mathcal{V}_2$) are again strictly decreasing (resp.\ increasing) in $\g$, see Appendix~\ref{PPVarDyn}. Thus, continuity and limit behavior of expectiles (see Theorem \ref{PropExp}, Properties \ref{MonoExp} and \ref{LimitExp}) allow us to straightforwardly characterize the existence and uniqueness of an optimal solution under all three premium principles via two boundary conditions. $\g^*$ can be calculated by (numerically) solving (\ref{CritPPESD}) resp. (\ref{CritPPVar}), see Algorithm~\ref{AlgoOptim}.

\begin{theorem}[Existence and uniqueness of $\a^*$ - Pure parametric insurance]\label{ExistencePP}
A unique solution $\a^*$ of (\ref{EqOpt}) exists under pure parametric insurance with:
\begin{itemize}
\item Expected value / standard deviation premium principle with loading $\rho > 0$ if and only if
\begin{equation}\label{LB_PP_ESD}
\frac{
\E\left[u'\bigl(w_0 - S + \bigl(1 - c^\circ \bigr)\essinf(S|\t \in \T)\bigr) \middle| \t \in \T \right]
}{
\E\left[u'\bigl(w_0 - S - c^\circ \essinf(S|\t \in \T)\bigr) \middle| \t \notin \T \right]
} > 
b^\circ
\end{equation}
as well as
\begin{equation}\label{UB_PP_ESD}
\frac{
\E\left[u'\left(w_0 - S + \bigl(1 - c^\circ \bigr) k\right) \middle| \t \in \T \right]
}{
\E\left[u'\bigl(w_0 - S - c^\circ k\bigr) \middle| \t \notin \T \right]
} < 
b^\circ
\end{equation}
for some $k  \in \bigl( \essinf(S|\t \in \T) , \, \esssup(S|\t \in \T) \bigr)$, $\circ \in \{\text{E},\text{SD}\}$, and
\begin{equation*}
b^\circ := \frac{\P(\t \notin \T) \, c^\circ}{\P(\t \in \T) \,\bigl(1 - c^\circ\bigr)} =\begin{cases}
1 + \frac{\rho}{1 - (1 + \rho)\P(\t \in \T)}, & \circ = \text{E}, \\
1 + \frac{\rho}{\sqrt{\P(\t \in \T)\P(\t \notin \T)} - \rho \, \P(\t \in \T)}, & \circ = \text{SD}.
\end{cases} 
\end{equation*}
\item Variance premium principle with loading $\rho > 0$ if and only if 
\begin{equation}
\label{LB_PP_Var}
\frac{\tilde{c}\bigl(\essinf(S|\t \in \T)\bigr)
\E\left[u'\left(w_0 - S + \essinf(S|\t \in \T) - \pi_0 \right)| \t \in \T \right]
}{
\E\left[u'\bigl(w_0 - S - \pi_0 \bigr) | \t \notin \T \right]
} > 
b^\text{V},
\end{equation}
as well as
\begin{equation}
\label{UB_PP_Var}
\frac{
\tilde{c}(k)\E\left[u'\left(w_0 - S +  \bigl(1 - \tilde{R}(k) \bigr)  k \right)| \t \in \T \right]
}{
\E\left[u'\bigl(w_0 - S - \tilde{R}(k)  k\bigr) | \t \notin \T \right]
} < 
b^\text{V},
\end{equation}
for some $k  \in \bigl( \essinf(S|\t \in \T) , \, \esssup(S|\t \in \T) \bigr)$. Here,
\begin{gather*}
\tilde{R}(x) := \P(\t \in \T) \bigl[1 + \rho x \P(\t \notin \T) \bigr], \qquad  \tilde{c}(x) := \frac{1}{\P(\t \in \T)\bigl[1 +2 \rho x \P(\t \notin \T)\bigr]} - 1,\\
\pi_0 := \tilde{R}\bigl(\essinf(S|\t \in \T) \bigr) \essinf(S|\t \in \T), \qquad b^\text{V} := \frac{\P(\t \notin \T)}{\P(\t \in \T)}.
\end{gather*}
\end{itemize} 
\end{theorem}

\begin{algorithm}
\caption{Derivation of $\gamma^*$} \label{AlgoOptim}
\textbf{Input:} 
\begin{itemize} 
\item Sufficiently large sample\footnote{Historical or simulated} of loss/index observations $(S_i,\theta_i)_{i = 1,\ldots,N}$
\item Auxiliary variables $u(\cdot)$, $w_0$, $\rho$
\end{itemize}
\textbf{Do:}
\begin{enumerate}
\item Split $(S_i,\theta_i)_{i = 1,\ldots,N}$ into $(S^{(1)}_i,\theta^{(1)}_i)$ and $(S^{(2)}_i,\theta^{(2)}_i)$ s.t. $\theta^{(j)}_i \in \T \Leftrightarrow j = 1$.
\item Approximate $\g \mapsto \V_j(\g)$ by its sample version $\hat{\V}_j(\g)$ on $(S^{(j)}_i,\theta^{(j)}_i)$, $j=1,2$.
\item Find $\g^*$ s.t. $\hat{\V}_1(\g^*) = \hat{\V}_2(\g^*)$.
\item Set $\a^* = \frac{1}{2}\mathds{1}_{\{\g^* = \frac{1}{2}\}} + \frac{\g^* - \sqrt{\g^* - (\g^*)^2}}{2\g^* - 1}\mathds{1}_{\{\g^* \neq \frac{1}{2}\}}$.
\end{enumerate}
\end{algorithm}

\begin{remark}[Optimal insurance choice under violated boundary conditions for pure parametric contracts]\label{PPVioBound}
It is immediate from the behaviour of $\mathcal{V}_1$ and $\mathcal{V}_2$ that at most one of the conditions in Theorem \ref{ExistencePP} can be violated under a given premium principle $\pi^\circ_\g$, $\circ \in \{\text{E},\text{SD},\text{V}\}$. While no $\alpha\in (0,1)$ will be optimal in this case, we can still make some inference on the policyholder's optimal insurance decision (see Appendix \ref{PP_VioB_proof} for detailed derivations):
\begin{itemize}
\item Lower boundary condition (\ref{LB_PP_ESD}), resp. (\ref{LB_PP_Var}), violated:\\
In this case, $\U(\gamma)$ is strictly decreasing on $(0,1)$. Thus, the policyholder will either choose the smallest possible $\alpha$\footnote{As there may not be an insurance contract available for values of $\alpha$ close to zero.} or opt for no insurance at all, depending on whether
\begin{equation*}
\min_\a \, \U \bigl( \gamma(\a) \bigr) > \U_0 := \E[u(w_0 - S)]
\end{equation*}
or not. A sufficient condition for preference for no insurance is given by
\begin{equation}\label{LB_noInsurance_EVSD}
\mathcal{V}_0 := \frac{
\E[u'(w_0 - S) | \t \in \T]
}{
\E[u'(w_0 - S) | \t \notin \T]
} \le 
 b^\circ, \quad \circ \in \{\text{E},\text{SD}\}
\end{equation}
under the expected value / standard deviation principle and by
\begin{equation}\label{LB_noInsurance_Var}
\mathcal{V}_0 \le 1
\end{equation}
under the variance principle. Note that (\ref{LB_noInsurance_EVSD}) and (\ref{LB_noInsurance_Var}) are always fulfilled if (\ref{LB_PP_ESD}) resp. (\ref{LB_PP_Var}) are violated and $\essinf(S|\t \in \T) = 0$.
\item Upper boundary condition (\ref{UB_PP_ESD}), resp. (\ref{UB_PP_Var}), violated:\\
This implies that $\U(\gamma)$ is strictly increasing on $(0,1)$, resulting in the policyholder choosing $\a$ as large as possible. Recalling the interpretation of $\a$ in the context of Definition~\ref{DefBROpt}, this would mean that the policyholder is always willing to accept higher premiums and more positive basis risk for a decrease in negative basis risk. Naturally, one would expect such a policyholder to prefer a full coverage indemnity insurance contract that completely eliminates basis risk. Indeed, if one considers such an alternative contract using the same premium principle with loading $\rho_\text{I}$, a sufficient condition for the policyholder to prefer the indemnity coverage over any pure parametric contract with basis risk weighting $\a \in (0,1)$ is given by
\begin{alignat*}{2}
\esssup(S|\t \in \T) & > \frac{\rho_\text{I}}{\rho} \frac{\E[S]}{\P(\t \in \T)} ,  && \qquad (\text{Expected} \ \text{value} \ \text{principle})\\
\esssup(S|\t \in \T) & > \left(\frac{\rho_\text{I}}{\rho}\right)^2  \frac{\Var(S)}{\Var\left(\mathds{1}_{\{\t \in \T\}} \right)} , && \qquad (\text{Standard} \ \text{deviation} \ \text{principle})\\
\bigl(\esssup(S|\t \in \T)\bigr)^2 &> \frac{\rho_\text{I}}{\rho}  \frac{\Var(S)}{\Var\left(\mathds{1}_{\{\t \in \T\}} \right)}. && \qquad (\text{Variance} \ \text{principle})
\end{alignat*}
Especially, note that these conditions are always fulfilled if $\esssup(S|\t \in \T) = \infty$. 
\end{itemize}
\end{remark}

\begin{remark}[Adapting boundary conditions for restriction to subintervals]\label{DiffBound}
We can adapt Theorem \ref{ExistencePP} to characterize the optimal $\alpha^*$ if only parametric contracts with $\g \in [\ubar{\g},\bar{\g}] \subsetneq (0,1)$ should be investigated. Here, we replace $\essinf(S|\t \in \T)$ and $\esssup(S|\t \in \T)$ by $e_{\ubar{\g}}(S|\t \in \T)$, resp. $e_{\bar{\g}}(S|\t \in \T)$. If both conditions are fulfilled, there exists a unique $\a^* \in (\ubar{\g},\bar{\g})$. Else, $\a^*$ is uniquely given by either $\ubar{\g}$ (if (\ref{LB_PP_ESD}), resp. (\ref{LB_PP_Var}), is violated) or by $\bar{\g}$ (if (\ref{UB_PP_ESD}), resp. (\ref{UB_PP_Var}), is violated).\footnote{Especially, $\a^*$ can now always be defined, contrary to the non-compact case of investigating $\g \in (0,1)$.} Restrictions to the admissible levels $\g$ could, e.g., be justified if only ``realistic'' $\a \ge 0.5$\footnote{Ensuring negative basis risk is penalized at least as much as positive basis risk, see~\cite{Maier2025}.} are of interest or if a regulator bars the fixed payment from approaching $\esssup(S|\t \in \T)$ arbitrarily close to avoid overinsurance. 
\end{remark}

As an example, we derive the boundary conditions as well as an explicit expression for $\a^*$ under exponential utility with expected value premium principle and visualize different scenarios detailed in Remark \Ref{PPVioBound}. Here, (\ref{EU_LB}) and (\ref{EU_UB}) follow directly from Theorem \ref{ExistencePP}. $e_{\g^*}(S| \t \in \T)$ can easily be derived by rearranging $\mathcal{V}_1(\g^*) = \mathcal{V}_2(\g^*)$. From this, the optimal weighting (\ref{OptEU}) is extracted by using (\ref{ExpSolve}) and (\ref{alphagamma}). Especially, note that - in line with standard results under exponential utility - the policyholder's initial wealth $w_0$ does not influence the optimal basis risk weighting $\a^*$.

\begin{corollary}[Optimal basis risk weighting under exponential utility and expected value principle]\label{CorExp}
Assume that $u$ is of the form 
\begin{equation*}
u(x) = 1 - \exp(-\beta x), \; \beta > 0. 
\end{equation*}
Then, the unique solution $\a^*_\text{exp}$ to (\ref{EqOpt}) under the expected value principle exists if and only if
\begin{align}
\frac{
\E \left[\exp(\beta   S) \middle| \t \in \T \right]
}{
\E\left[\exp(\beta   S) \middle| \t \notin \T \right]
} \cdot \exp\bigl(-\beta\, \essinf(S|\t \in \T)\bigr) > 
1 + \frac{\rho}{1 - (1 + \rho)\P(\t \in \T)}, \label{EU_LB} \\
\frac{
\E \left[\exp(\beta  S) \middle| \t \in \T \right]
}{
\E\left[\exp(\beta  S) \middle| \t \notin \T \right]
} \cdot \exp(-\beta  k) < 
1 + \frac{\rho}{1 - (1 + \rho)\P(\t \in \T)}, \label{EU_UB}
\end{align}
for some $k  \in \bigl( \essinf(S|\t \in \T) , \, \esssup(S|\t \in \T) \bigr)$ and is given by
\begin{align}
\a^*_\text{exp} &= \a \left( \frac{L_\T(x_\text{exp}) - x_\text{exp}F_\T(x_\text{exp})}{2\bigl(L_\T(x_\text{exp}) - x_\text{exp}F_\T(x_\text{exp})\bigr) + x_\text{exp} - \E[S|\t \in \T]} \right), \label{OptEU}\\
x_\text{exp} &:= e_{\g^*}(S| \t \in \T) = -\frac{1}{\beta} \left\{ \log\left( \frac{c^\text{E}}{1 - c^\text{E}}\right) + \log \left( \frac{\E\left[\exp(\beta S) \Ind_{\{\t \notin \T\}}\right]}{\E\left[\exp(\beta S) \Ind_{\{\t \in \T\}}\right]} \right) \right\}, \nonumber \\
F_\T(s) &:= \P(S \le s | \t \in \T), \quad L_\T(x) := \int_{-\infty}^x s \, \dF_\T(s). \nonumber
\end{align}
\end{corollary}

\begin{table}[t]
\begin{center}
\begin{tabular}{c|ccc|ccc}
\toprule
$\bm{\ell}$ & $\bm{\rho^{(\ell)}}$ &  $\bm{a^{(\ell)}}$ & $\bm{ \P\left( \t \in \T^{(\ell)}\right) }$ & \textbf{(\ref{LB_noInsurance_EVSD}) violated} & \textbf{(\ref{EU_UB}) violated} & \textbf{``No insurance''} \\
\midrule
1 & 0.1 & 4 & 0.5 & \cmark & \xmark & \xmark \\
\midrule
2 & 0.2 & 4 & 0.5 & \cmark & \xmark & \cmark \\
\midrule
3 & 0.3 & 1 & 0.6 & \xmark & \xmark & \cmark \\ 
\bottomrule
\end{tabular}
\caption{Individual parameters and optimal insurance decision for the policyholders $\ell=1,2,3$ in Example \ref{ExpUPPExample} to visualize the circumstances described in Remark~\ref{PPVioBound}. The column "No insurance" indicates whether the policyholder prefers being uninsured over any parametric contract. Details on why (\ref{EU_UB}) is never violated in this setting are provided at the end of Example~\ref{ExpUPPExample}.}
\label{ExU_Table}
\end{center}
\end{table}

\begin{example}[Exponential utility]\label{ExpUPPExample}
We investigate the optimal insurance decision for three policyholders $(\ell=1,2,3)$ with equal initial wealth $w_0 = 10$, exponential utility
\begin{equation*}
u(x) = 1 - \exp(-\beta x), \quad \beta = 0.1,
\end{equation*}
and conditional loss
\begin{align*}
&\P\left(S^{(\ell)} = 5 \, \middle | \, \t \in \T^{(\ell)}\right) = \frac{1}{2} = \P\left(S^{(\ell)} = 10 \, \middle | \, \t \in \T^{(\ell)}\right), \\
&\P\left(S^{(\ell)} = 0 \, \middle| \, \t \notin \T^{(\ell)}\right) = \frac{1}{2} = \P\left(S^{(\ell)} = a^{(\ell)}\, \middle | \, \t \notin \T^{(\ell)}\right), \quad a^{(\ell)} \in \R_{>0},
\end{align*}
for a pure parametric contract with index $\t$ and expected value premium principle.\footnote{Note that \cite[Example 2.1]{Daouia2024} and linearity of expectiles yield $e_\g(S|\t \in \T) = 5(1+\g)$, ensuring the differentiability of $Y^*_\g$.} The policyholders differ w.r.t. their trigger area $\T^{(\ell)}$, their premium loading $\rho^{(\ell)}$, and their maximum possible loss $a^{(\ell)}$ if no payment is triggered, as detailed in Table \ref{ExU_Table}. 

For all policyholders, condition (\ref{EU_LB}) is violated, resulting in the following optimal insurance decision:
\begin{itemize}
\item For policyholder 1, the alternative condition (\ref{LB_noInsurance_EVSD}) is violated, as $\mathcal{V}^{(1)}_0 \approx 1.753 > 1.222 \approx b^{(1),\text{E}}$. Since
\begin{align*}
\E[u(w_0 - S)] \approx 0.369< 0.378 &\approx \E\Bigl[u\Bigl(w_0 - S^{(1)} + \essinf\bigl(S^{(1)} \big| \,\t \in \T\bigr) (\Ind_{\{\t \in \T\}} - c^{(1),\text{E}})\Bigr)\Bigr] \\
&= \lim_{\g \searrow 0} \E\bigl[u\bigl(w_0 - S^{(1)} + Y^{(1),*}_\g - \pi^{(1),\text{E}}_\g\bigr)\bigr],
\end{align*}
policyholder 1 will choose the smallest possible $\a^{(1)}$, provided it is close enough to $0$, over no insurance.
\item For policyholder 2, the alternative condition (\ref{LB_noInsurance_EVSD}) is again violated, as $\mathcal{V}^{(2)}_0 \approx 1.753 > 1.5=b^{(2),\text{E}}$. Since
\begin{align*}
\E[u(w_0 - S)] \approx 0.369> 0.362 &\approx \E\Bigl[u\Bigl(w_0 - S^{(2)} + \essinf\bigl(S^{(2)} \big| \,\t \in \T\bigr) (\Ind_{\{\t \in \T\}} - c^{\text{E}})\Bigr)\Bigr] \\
&= \lim_{\g \searrow 0} \E\bigl[u\bigl(w_0 - S^{(2)} + Y^{(2),*}_\g - \pi^{(2),\text{E}}_\g\bigr)\bigr],
\end{align*}
policyholder 2 prefers no insurance over any pure parametric contract.
\item For policyholder 3, the alternative condition (\ref{LB_noInsurance_EVSD}) holds, as $\mathcal{V}^{(3)}_0 \approx 2.074 \le 2.364 \approx b^{(3),\text{E}}$. Hence, they prefer no insurance over any pure parametric contract.
\end{itemize}

Note that in the setting of Corollary \ref{CorExp}, we have
\begin{gather*} 
0 \le \E\bigl[\exp\bigl(\beta (S - k) \bigr) \big| \, \t \in \T \bigr] \le \E\bigl[\exp\bigl(\beta (S - k)_+\bigr)  \big| \, \t \in \T  \bigr] \underset{k \nearrow \esssup(S|\t \in \T)}{\longrightarrow} 1, \\
\E\bigl[ \exp(\beta S ) | \, \t \notin \T \bigr]  \ge 1, \qquad b^\text{E} > 1,
\end{gather*}
for any $S \ge 0$ a.s. and $\rho > 0$. Thus, condition (\ref{EU_UB}) can never be violated under the expected value premium principle.
\end{example}

\subsection{Parametric index insurance}\label{PI}

We now extend our considerations to the more general payment schemes (\ref{PS_PI}) of parametric index insurance. Here, our derivatives of interest are given by
\begin{align*}
\U'_1(\g) &= \int_\T \bigl(e'_\g(S|\theta) - \pi'_\g\bigr) \int u'\bigl(w_0 - s + e_{\g}(S|\theta) -\pi_{\g} \bigr) \, \dF_{\theta}(s) \, \dG(\theta),\\
\U'_2(\g) &= - \int_{\T^c}  \pi'_\g  \int u'\bigl(w_0 - s -\pi_{\g} \bigr) \, \dF_{\theta}(s) \, \dG(\theta).
\end{align*}

In contrast to Section \ref{PP}, the basis risk-optimal payout and its derivative(s) on $\T$ now depend on the realization of $\t$. Thus, general assessment of the optimal $\a^*$ is made difficult by the fact that - to the best of our knowledge - no general dominance results for (conditional) expectiles' derivatives of order $\ge 2$ exist. However, if we assume that the behaviour of $e'_\g(S|\t)$ w.r.t. $\t$ and $\g$ can be represented as arising from two separate effects, we can handily extend our observations from Section \ref{PP} to the domain of parametric index insurance. While Assumption~\ref{SepDerivAssum} is considered to hold for the rest of this section, we stress that it is not a \emph{necessary} condition for the existence of (a unique) $\a^*$. Some numerical evidence is provided in Appendix~\ref{PINoAssum}.

\begin{assumption}[Seperable expectile derivatives]\label{SepDerivAssum}
There exist functions $h_1:\T \to \R$ and \mbox{$h_2:(0,1) \to \R$} such that
\begin{equation*}
e'_\g(S|\theta) = h_1(\theta)h_2(\g) , \;  \forall \, \g \in (0,1), \, \theta \in \T.
\end{equation*}
By Property \ref{MonoExp} of Theorem \ref{PropExp}, we can w.l.o.g. assume $h_1$ and $h_2$ to be strictly positive.
\end{assumption}

The results of \cite{Daouia2024} allow us to recognize the exponential distribution as a potential family of distributions $(S|\t = \theta)_{\theta \in \T}$ that conforms to Assumption \ref{SepDerivAssum}. 

\begin{example}[Separable expectile derivatives - Exponential distribution]\label{SepExpDistr}
Assume that \mbox{$(S|\t = \theta) \sim \text{Exp}\bigl(1/\lambda(\theta)\bigr)$} with $\lambda: \T \to (a,b)$, $0< a < b$. Then, a.s.,    
\begin{equation*}
e_\g(S|\theta) = \lambda(\theta) \left[ 1 + W\left(\frac{2\g - 1}{1 - \g}\exp(-1) \right) \right]  , \;  \forall \, \g \in (0,1), \, \theta \in \T,
\end{equation*}
where $W(\cdot)$ denotes the principal branch of the Lambert $W$ function.\footnote{More details including an analytic expression for $W'$ are provided in \cite{Corless1996}.} Especially, $(S,\t)$ fulfills Assumption \ref{SepDerivAssum} with
\begin{equation*}
h_1(\theta) = \lambda(\theta), \quad h_2(\g)=\frac{\exp(-1)}{(1 - \g)^2} W'\left(\frac{2\g - 1}{1 - \g} \exp(-1) \right).
\end{equation*}
\end{example}

The following proposition characterizes separable expectile derivatives and shows that it is sufficient and (under a rather mild additional condition, see Remark \ref{SepDerivRemark}) necessary that $(S|\t = \theta)_{\theta \in \T}$ belongs to a location-scale family for Assumption~\ref{SepDerivAssum} to hold. A proof is given in Appendix~\ref{ProofExpDeriv}. Example~\ref{LocScaleFamily} lists some potential families of severity distributions, including the Exponential distribution already discussed in Example~\ref{SepExpDistr}. We would especially like to stress that Assumption~\ref{SepDerivAssum} is fulfilled if the distribution of $S|\t$ can be represented through a regression approach with additive or multiplicative errors.\footnote{An example for this is presented by the model (\ref{Model_Loss}) used in our simulation study in Section~\ref{SimStudy}.}

\begin{proposition}[Characterization of separable expectile derivatives]\label{SepDerivProp}
It holds:
\begin{equation}\label{proof1}
e'_\g(S|\t) = h_1(\t)h_2(\g) \quad \Leftrightarrow \quad e_\g(S|\t) = h_1(\t)H_2(\g) + H_3(\t)
\end{equation}
for appropriate $H_2: (0,1) \to \R$ with $H'_2(\g) = h_2(\g)$ and $H_3:\T \to \R$. \\
Further, if $h_1(\t) \in \L^\infty(\t)$, 
\begin{equation}\label{proof2}
e'_\g(S|\t) = h_1(\t)h_2(\g) \quad \Leftrightarrow \quad S|\t =_{d} \mu(\t) + \sigma(\t)\cdot Z
\end{equation}
for appropriate $\mu: \T \to \R$, $\sigma: \T \to \R_{>0}$, and $Z \in \L^2$ with $\mu(\t) \in \L^2(\t)$, $\sigma(\t) \in \L^\infty(\t)$, $Z \perp \t$.
\end{proposition}

\begin{remark}[A sufficient condition for $h_1(\t) \in \L^\infty(\t)$]\label{SepDerivRemark}
For any $\theta \in \T$, we can write
\begin{equation*}
h_1(\theta)h_2\left(\frac{1}{2}\right) = e'_{0.5}(S|\theta) \underset{(*)}{=} 2\,\E\Bigl[\bigl| S - \E[S|\t = \theta] \bigr| \Big| \t = \vartheta \Bigr] \le 2\bigl(1 + \Var(S|\t = \theta)\bigr), 
\end{equation*}
where $(*)$ follows from Equation (\ref{AltExpDerivative}). Thus, $\underset{\theta \in \T}{\esssup}\bigl(\Var(S|\t = \theta)\bigr)<\infty$ implies $h_1(\t) \in \L^\infty(\t)$.
\end{remark}

\begin{example}[Location-scale families for loss severity]\label{LocScaleFamily}
The following are examples for non-negative location-scale families of distributions:
\begin{itemize}
\item \emph{Exponential} distribution $\text{Exp}(\lambda)$: $F_\lambda(x) = 1 - \exp(-\lambda x)$, $x>0$, $\lambda > 0$,
\begin{equation*}
\text{Exp}(\lambda) =_d \frac{1}{\lambda}Z, \quad Z \sim \text{Exp}(1).
\end{equation*}
\item \emph{Gamma} distribution $\text{Gamma}(r, \kappa)$: $F_{r,\kappa}(x) = \frac{1}{\Gamma(r)}\int_0^{\kappa x} t^{r - 1} \exp(-t) \, dt$, $x>0$, $r,\kappa > 0$,
\begin{equation*}
\text{Gamma}(r, \kappa) =_d \frac{1}{\kappa} Z, \quad Z \sim \text{Gamma}(r, 1).
\end{equation*}
\item \emph{Log-normal} distribution $\mathcal{LN}(\mu, \sigma^2)$: $F_{\mu, \sigma}(x) = \Phi\left(\frac{\log(x) - \mu}{\sigma}\right)$, $x>0$, with $\Phi(\cdot)$ the cdf of $\mathcal{N}(0,1)$, $\mu \in \R$, $\sigma > 0$,
\begin{equation*}
\mathcal{LN}(\mu, \sigma^2) =_d \exp(\mu) \cdot Z, \quad Z \sim \mathcal{LN}(0, \sigma^2).
\end{equation*}
\item \emph{Pareto (type II)} distribution $P_\text{II}(\mu,\sigma,\beta)$: $F_{\mu,\sigma,\beta}(x) = 1 - \left( 1 + \frac{x - \mu}{\sigma}\right)^{-\beta}$, $x \ge \mu \in \R_{>0}$, $\sigma,\beta > 0$,
\begin{equation*}
P_\text{II}(\mu,\sigma,\beta) =_d \mu + \sigma Z, \quad Z \sim P_\text{II}(0,1,\beta).
\end{equation*}
\end{itemize} 
\end{example}

We can now generalize the results of Theorem \ref{ExistencePP} for the expected value and variance premium principle to the corresponding parametric index insurance contracts. As the derivation of results under the standard deviation principle would involve differentiating $h_2(\g)$ and thus require (at the very least) additional assumptions, we exclude this case from further consideration. 

Starting with the expected value principle, i.e.,
\begin{equation*} 
\pi^E_\g = (1+\rho)\int_\T e_\g(S|\theta) \, \dG(\theta) = (1+\rho)\int_\T [h_1(\theta)H_2(\g) + H_3(\theta)] \, \dG(\theta),
\end{equation*}
Assumption \ref{SepDerivAssum} yields
\begin{align*}
\U'_1(\g) &= h_2(\g) \int_\T  \Delta_1(\theta) \int u'\bigl(w_0 - s + \Delta_1(\theta) H_2(\g) + \Delta_3(\theta) \bigr) \, \dF_{\theta}(s) \, \dG(\theta),\\
\U'_2(\g) &= - h_2(\g) \left[ (1+\rho)\int_\T h_1(\vartheta) \, \dG(\vartheta)\right] \int_{\T^c} \int u'\left(w_0 - s - \pi_\g^E \right) \, \dF_{\theta}(s) \, \dG(\theta),
\end{align*}
with
\begin{align*}
\Delta_1(\theta) := h_1(\theta) - (1+\rho)\int_\T h_1(\vartheta) \, \dG(\vartheta), \quad \Delta_3(\theta) := H_3(\theta) - (1+\rho)\int_\T H_3(\vartheta) \, \dG(\vartheta).
\end{align*}
Paralleling our prior derivations, we investigate the behaviour of $\V_i(\g) := (-1)^{i-1} \U'_i(\g)/h_2(\g)$, $i=1,2$, for which we find
\begin{align*}
\V_1'(\g) &:= h_2(\g) \int_\T  \Delta_1^2(\theta) \int u''\bigl(w_0 - s + \Delta_1(\theta) H_2(\g) + \Delta_3(\theta) \bigr) \, \dF_{\theta}(s) \, \dG(\theta) \le 0, \\
\V_2'(\g) &:= - h_2(\g) \left[ (1+\rho)\int_\T h_1(\vartheta) \, \dG(\vartheta)\right]^2 \int_{\T^c} \int u''\left(w_0 - s - \pi_\g^E \right) \, \dF_{\theta}(s) \, \dG(\theta) > 0.
\end{align*}
When premiums are computed with the variance principle,
\begin{align}\label{PremiumPIVar}
\pi^V_\g &=  \int_\T \bigl[ h_1(\theta)H_2(\g) +  H_3(\theta)\bigr] \,  \dG (\theta)+ \rho \bigl[H_2^2(\g) V_1 + 2 H_2(\g) V_{13} + V_3 \bigr], \\
(\pi^V_\g)' &= h_2(\g) \underbrace{\left[\int_\T h_1(\theta) \,  \dG (\theta) + 2\rho H_2(\g)V_1 + 2\rho V_{13}\right]}_{=:R(\g)}, \nonumber
\end{align}
where
\begin{align*}
V_1 &:= \Var\left( h_1(\t)\Ind_{\{\t \in \T\}}\right) = \int \left( h_1(\theta)\Ind_{\{\theta \in \T\}} -  \int_\T h_1(\vartheta) \, \dG(\vartheta) \right)^2 \dG(\theta),\\
V_3 &:=  \Var\left( H_3(\t)\Ind_{\{\t \in \T\}}\right) = \int \left( H_3(\theta)\Ind_{\{\theta \in \T\}} -  \int_\T H_3(\vartheta) \, \dG(\vartheta) \right)^2 \dG(\theta),\\
V_{13} &:= \Cov\left( h_1(\t)\Ind_{\{\t \in \T\}}, H_3(\t)\Ind_{\{\t \in \T\}} \right) \\
&\, = \int \left( h_1(\theta) H_3(\theta)\Ind_{\{\theta \in \T\}} -  \int_\T h_1(\vartheta) \, \dG(\vartheta) \cdot \int_\T H_3(\vartheta) \, \dG(\vartheta) \right) \dG(\theta).
\end{align*}
Analogously to above, we derive
\begin{align*}
\U'_1(\g) &= h_2(\g) \int_\T  \bigl[h_1(\theta) - R(\g) \bigr] \int u'\bigl(w_0 - s + e_\g(S|\theta) - \pi^V_\g \bigr) \, \dF_{\theta}(s) \, \dG(\theta),\\
\U'_2(\g) &=- h_2(\g) \int_{\T^c}   R(\g)  \int u'\bigl(w_0 - s - \pi^V_\g \bigr) \, \dF_{\theta}(s) \, \dG(\theta),\\
\V'_1(\g) &= h_2(\g) \int_\T  \bigl[h_1(\theta) - R(\g) \bigr]^2 \int u''\bigl(w_0 - s + e_\g(S|\theta) - \pi^V_\g \bigr) \, \dF_{\theta}(s) \, \dG(\theta)\\
&\hspace{4cm} - \int_\T  R'(\g) \int u'\bigl(w_0 - s + e_\g(S|\theta) - \pi^V_\g \bigr) \, \dF_{\theta}(s) \, \dG(\theta)  < 0,\\
\V'_2(\g) &= \int_{\T^c}  R'(\g)  \int u'\bigl(w_0 - s - \pi^V_\g \bigr) \, \dF_{\theta}(s) \, \dG(\theta) \\
& \hspace{4cm} - \int_{\T^c}  R^2(\g)  \int u''\bigl(w_0 - s - \pi^V_\g \bigr) \, \dF_{\theta}(s) \, \dG(\theta) > 0,\\
R'(\g) &= 2\rho h_2(\g) V_1 > 0, 
\end{align*}
Thus, we can again ensure existence and uniqueness of $\g^* \in (0,1)$ through two boundary conditions and derive it by (numerically) solving $\V_1(\g^*) = \V_2(\g^*)$, see Algorithm~\ref{AlgoOptim}. We formalize our results in Theorem~\ref{ExistencePI} and note that they match the conditions of Theorem \ref{ExistencePP} if one assumes $(S|\t = \theta) =_d (S|\t \in \T)$ for all $\theta \in \T$.\footnote{As the parametric index insurance contract is equal to the pure parametric contract in this setting, the agreement of Theorems \ref{ExistencePP} and \ref{ExistencePI} presents a natural sanity check.} For notational convenience, we set
\begin{equation*}
H_2(0) := \lim_{\g \searrow 0} H_2(\g) \in [0,\infty), \qquad H_2(1) := \lim_{\g \nearrow 1} H_2(\g) \in (0,\infty],
\end{equation*}
whose well-definedness is guaranteed by (\ref{proof1}). 

\begin{theorem}[Existence and uniqueness of $\a^*$ - Parametric index insurance]\label{ExistencePI}
Under Assumption~\ref{SepDerivAssum}, a unique solution $\a^*$ of (\ref{EqOpt}) exists under parametric index insurance with:
\begin{itemize}
\item Expected value premium principle with loading $\rho > 0$ if and only if
\begin{equation}\label{LB_PI_EV}
\frac{
\E\left[\Delta_1(\t)\, u'\bigl(w_0 - S + \Delta_1(\t) H_2(0)+ \Delta_3(\t) \bigr) \middle| \t \in \T \right]
}{
\E\left[u'\left(w_0 - S - (1 + \rho) \int_\T \bigl(h_1(\theta) H_2(0) + H_3(\theta)\bigr) \, \dG(\theta) \right) \middle| \t \notin \T \right]
} > 
b^\text{E}
\end{equation}
and
\begin{equation}\label{UB_PI_EV}
\frac{
\E\left[\Delta_1(\t) \, u'\bigl(w_0 - S + \Delta_1(\t) k+ \Delta_3(\t) \bigr) \middle| \t \in \T \right]
}{
\E\left[u'\left(w_0 - S - (1 + \rho) \int_\T \bigl(h_1(\theta) k + H_3(\theta)\bigr) \, \dG(\theta) \right) \middle| \t \notin \T \right]
} < 
b^\text{E}
\end{equation}
for some $H_2(0) < k < H_2(1)$ and
\begin{equation*}
b^\text{E} := \frac{(1 + \rho) \P(\t \notin \T)}{\P(\t \in \T)} \int_\T h_1(\theta) \, \dG(\theta).
\end{equation*}
\item Variance premium principle with loading $\rho > 0$ if and only if
\begin{equation}\label{LB_PI_V}
\hspace{-1cm}\frac{ 
\E\Bigl[ \Bigl(h_1(\t) - \tilde{R}\bigl(H_2(0)\bigr)\Bigr) \, u'\Bigl(w_0 - S + \essinf(S|\t) - \tilde{\pi}^V\bigl(H_2(0)\bigr) \Bigr) \Big| \t \in \T \Bigr]
}{
\E\left[\tilde{R}\bigl(H_2(0)\bigr) \, u'\left(w_0 - S - \tilde{\pi}^V\bigl(H_2(0)\bigr) \right) \middle| \t \notin \T \right]
} > \frac{\P(\t \notin \T)}{\P(\t \in \T)}
\end{equation}
and 
\begin{equation}\label{UB_PI_V}
\frac{
\E\Bigl[ \bigl(h_1(\t) - \tilde{R}(k)\bigr) \, u'\bigl(w_0 - S + h_1(\t)k + H_3(\t) - \tilde{\pi}^V(k) \bigr) \Big| \t \in \T \Bigr]
}{
\E\left[\tilde{R}(k) \, u'\bigl(w_0 - S - \tilde{\pi}^V(k) \bigr) \middle| \t \notin \T \right]
} < \frac{\P(\t \notin \T)}{\P(\t \in \T)}
\end{equation}
for some $H_2(0) < k < H_2(1)$ with
\begin{align*}
\tilde{R}(k) &= 2\rho kV_1 + 2\rho V_{13} + \int_\T h_1(\theta)\, \dG(\theta), \\
\tilde{\pi}^V(k) &=\int_\T \bigl(h_1(\theta) k + H_3(\theta) \bigr)\, \dG(\theta)+  \rho \bigl(k^2V_1 + 2 kV_{13} + V_3\bigr).
\end{align*}
\end{itemize}
\end{theorem}

In view of the structural similarities to pure parametric insurance, it is not surprising that we can directly translate our observations on maximizing $\U(\g)$ over $[\ubar{\g},\bar{\g}] \subsetneq (0,1)$ from Remark~\ref{DiffBound} to parametric index insurance by replacing $H_2(0)$ and $H_2(1)$ by $H_2(\ubar{\g})$ and $H_2(\bar{\g})$. Likewise, mirroring Remark~\ref{PPVioBound}, we can compare the expected utility of parametric index insurance under violated boundary conditions to the utility of no insurance resp. full indemnity coverage. Detailed derivations are given in Appendix \ref{PI_VioB_proof}.

\begin{remark}[Optimal insurance choice under violated boundary conditions for parametric index insurance]\label{PIVioBound}
If one of the conditions in Theorem \ref{ExistencePI} is violated, $\U(\g)$ is strictly decreasing ((\ref{LB_PI_EV}) or (\ref{UB_PI_EV})) resp.\ strictly increasing ((\ref{LB_PI_V}) or (\ref{UB_PI_V})) on $(0,1)$. In the former case, if $\essinf(S|\t=\theta) = 0$ for almost all $\theta \in \T$, the policyholder strictly prefers no insurance to any parametric solution. In the latter case, a sufficient condition for preferring full coverage indemnity insurance over any parametric contract is given by
\begin{alignat*}{2}
\E\bigl[\esssup(S|\t) \big|\t \in \T\bigr] & > \frac{\rho_\text{I}}{\rho} \frac{\E[S]}{\P(\t \in \T)} ,  && \qquad \textrm{(Expected value principle)}\\
\Var \left(\esssup(S|\t)\mathds{1}_{\{\t \in \T\}} \right) &> \frac{\rho_\text{I}}{\rho }\Var(S), && \qquad\textrm{(Variance principle)}
\end{alignat*}
where we set $\Var \left(\esssup(S|\t)\mathds{1}_{\{\t \in \T\}} \right) := \infty$ if $\esssup(S|\t = \theta)=\infty$ for some $\theta \in \T$.\footnote{Recalling (\ref{PremiumPIVar}), one can easily see that this implies $\lim_{\g \to 1} H_2(\g) =  \infty$ and thus $\esssup(S|\t = \theta) = \infty$ for almost all $\theta \in \T$.}
\end{remark}

\begin{remark}[Choosing $\a$ in practical applications]
For a given contract framework $(S,\t, \T,\rho, w_0, \pi^\circ)$, we can map any utility $u$ fulfilling the boundary conditions of Theorem~\ref{ExistencePP}, resp.~\ref{ExistencePI}, to the optimal $\a^*$. However, providing an explicit form for this map is only feasible in specific circumstances (compare Corollary~\ref{CorExp}). In practice, insurers can offer a menu of contracts across an appropriate range of $\a$, which may be identified through surveys on the impact of basis risk on insurance demand (see, e.g., \cite{Jensen2018}). Additionally, policyholders not adequately represented in the default offerings can customize their contracts thanks to the high flexibility of parametric solutions, see \cite{SwissRe2024}.
\end{remark}

\section{Simulation study: Cat-in-a-circle hurricane insurance}\label{SimStudy}

\subsection{Setup and model assumptions}\label{SimSetup}

To visualize the findings from Section~\ref{TheoOpt}, we take inspiration from \cite{SwissRe2020} and consider cat-in-a-circle parametric protection against hurricane damage to an insured building with value $v$. More precisely, the contract specifications are as follows:
\begin{itemize}
\item A hurricane passing within $r= 50$ km of the property's location at 28.39°~N 81.56°~W\footnote{As maintaining confidentiality is integral for our (fictitious) insurer, the interested reader may need to rely on one of the numerous GPS location finders available online to identify the policyholder.} is considered an incident.
\item For each incident, compensation is based on the one-minute maximum sustained wind speed $\t$, measured in knots (kn).
\item Non-zero payments are triggered for wind speeds corresponding to Category 2 and above on the Saffir--Simpson scale, i.e. the trigger area is given by $\T = [83,\infty)$.
\item Premiums are calculated using the expected value premium principle (\ref{EV_Premium}).
\end{itemize}

\begin{figure}[t]
\center
\includegraphics[width = 0.75\textwidth, keepaspectratio]{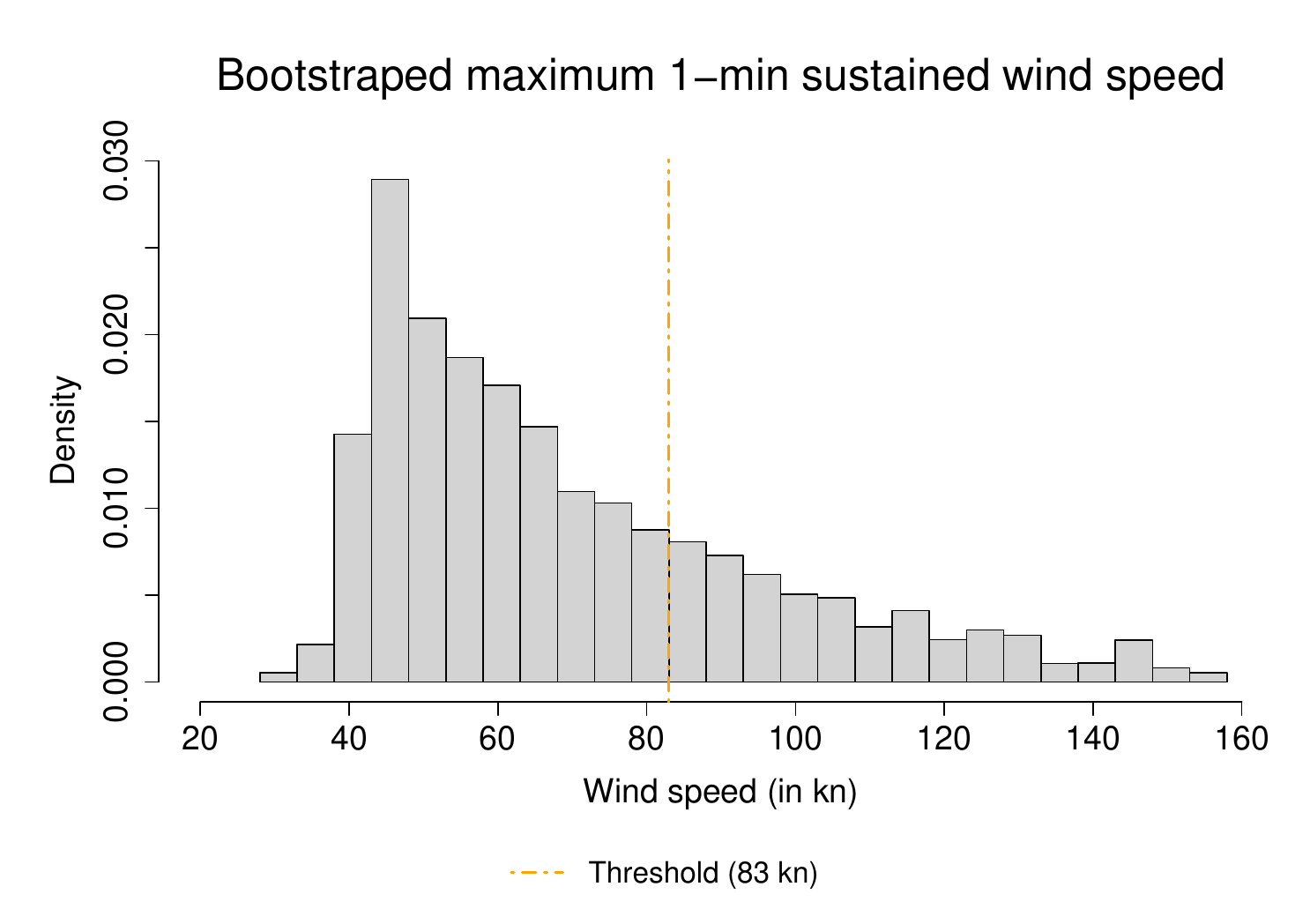}
\caption{Histogram of bootstrapped one-minute maximum sustained wind speed realizations $(\theta_i)_{i = 1,\ldots,n}$ for hurricanes passing the policyholder's trigger area and threshold above which a payout to the policyholder is triggered (orange line).}
\label{WindHisto}
\end{figure}

To approximate the expectations underlying our results in Theorems~\ref{ExistencePP} and~\ref{ExistencePI}, we generate $n = 10^6$ loss-index-realizations $(S_i,\theta_i)_{i = 1,\ldots,n}$. Hurricane wind speed modeling has evolved from using site-specific frameworks (see, e.g., \cite{Batts1980, Vickery1995}) to simulating from increasingly sophisticated hurricane track models (see, e.g., \cite{Vickery2000, Snaiki2020}). While we believe the latter approach to be a realistic candidate for the NatCat risk assessment process in the insurance industry, setting up a model for synthetic hurricane generation from scratch is out of this paper's scope. Thus, we rely on the results of Bloemendaal et al. \cite{Bloemendaal2020}, whose publicly available STORM data set \cite{BloemendaalData} contains $10\,000$ years of simulated hurricane activity. All tracks passing through the trigger area\footnote{Minimum distance to the circle's center was determined using the methodology described in \cite{Gade2010}, available in the Python package \texttt{nvector}.} were identified and the maximum wind speed\footnote{As STORM simulates 10-minute maximum sustained wind speeds, values were transformed back to 1-minute maxima using the methodology outlined in \cite{Bloemendaal2020}.} was recorded based on the measurements at all locations in and immediately before / after the trigger area. From these values, observations $(\theta_i)_{i = 1,\ldots,n}$ were simulated using non-parametric bootstrap, see~\cite{Efron1979,Efron1993}. The corresponding histogram is displayed in Figure~\ref{WindHisto}.

To derive the associated loss realizations $S_i$, we take inspiration from the hazard functions visualized in \cite[Figure 7]{Snaiki2023}. Thus, we assume the mean conditional losses $\mu(\t)$ to follow an S-shaped curve and include randomness through an appropriately scaled error term, setting
\begin{gather}
(S_i|\t = \theta_i) = \mu(\theta_i) + \sigma(\theta_i)\left[ \min \left\{\frac{p + q}{p},\frac{p + q}{q} \right\}\cdot  Z_i - \min \left\{1, \frac{p}{q} \right\}\right] , \quad p,q > 0, \; i=1,\ldots,n, \nonumber \\
\mu(\theta) = v\frac{1 - e^{-0.09 (\theta - 64)}}{1 + 150e^{-0.09 (\theta - 64)}}\mathds{1}_{\{\theta \ge 64 \}},\quad \sigma(\theta) =  \mu(\theta)\left[ 1 - \frac{\mu(\theta)}{v} \right],\quad Z_i \sim \mathrm{Beta}(p,q) \; \textrm{i.i.d.},   \label{Model_Loss}
\end{gather}

\begin{figure}[t]
\center
\includegraphics[width = \textwidth, keepaspectratio]{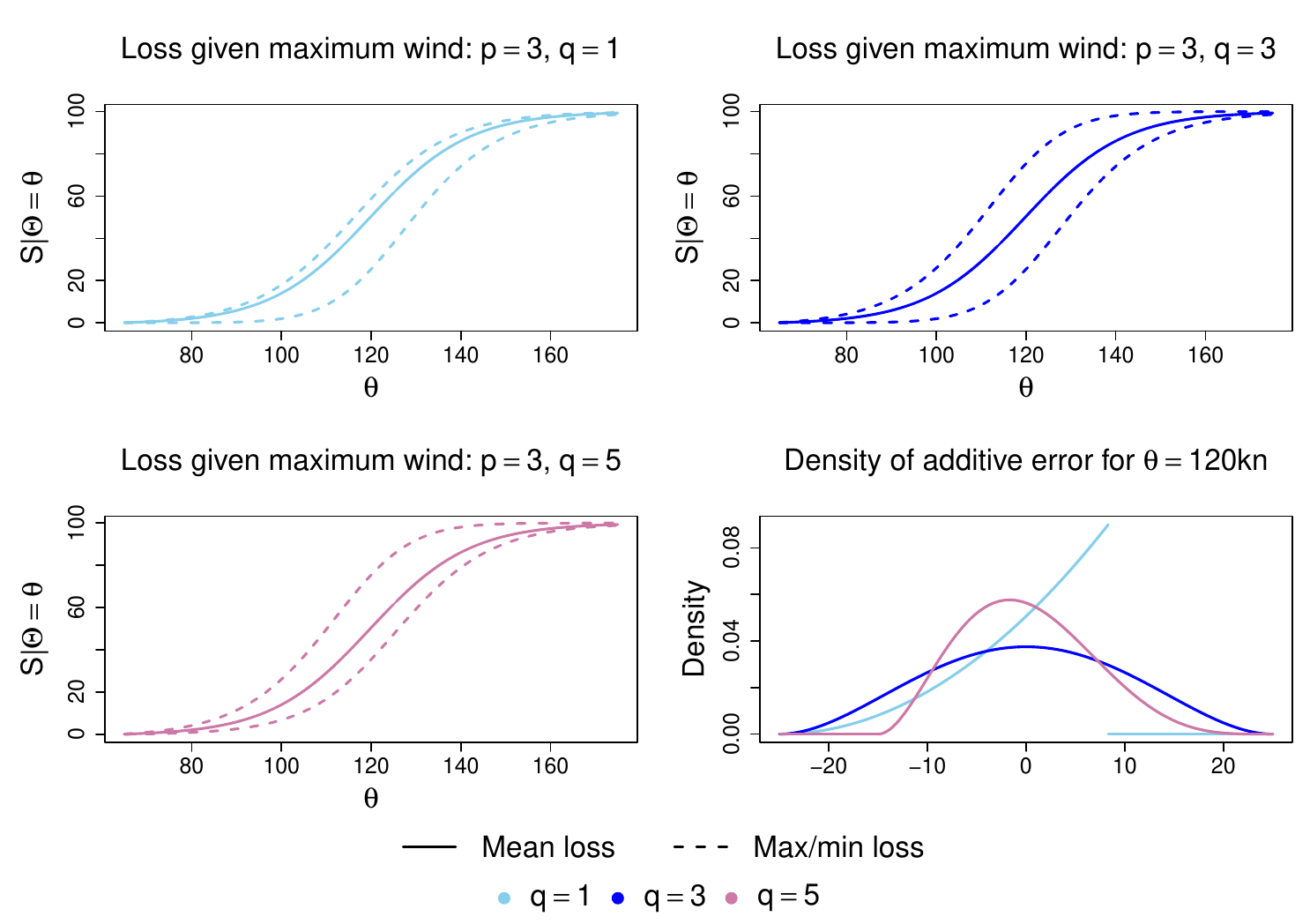}
\caption{Mean, maximum and minimum conditional loss $S|\t$  and densities of the error terms $ \sigma(\theta)\left[ \min \left\{\frac{p + q}{p},\frac{p + q}{q} \right\}\cdot  Z - \min \left\{1, \frac{p}{q} \right\}\right]$ for $\theta =120\textrm{kn}$ (bottom right) under different parameter choices ($p=3$, $q \in \{1,3,5\}$).}
\label{Loss_Error_Distributions}
\end{figure}

The choice of the additive error term in (\ref{Model_Loss}) is motivated by the following properties:
\begin{itemize}
\item Conditional losses always abide their natural boundaries, i.e. $S|\t \in [0,v]$ almost surely.
\item Conditional losses are less variable when their mean is close to the extremes $\{0,v\}$.
\item Errors can exhibit varying (a-)symmetry depending on the choices of $p$ and $q$ while leaving the conditional mean unaffected, see Figure \ref{Loss_Error_Distributions}.
\end{itemize}
Nonetheless, let us stress that our primary goal is to enjoy an educational example and we do not claim the above to be an accurate wind-loss model.

\subsection{Simulation results}
In the following, we shortly summarize the results of our simulations. Unless specified otherwise, all figures and observations refer to a policyholder with exponential utility $u(x) = 1 - \exp(-\beta x)$, risk aversion parameter $\beta = 0.15$, premium loading $\rho = 0.2$, and building value $v = 100$ (i.e. all monetary amounts are expressed in percentage points of $v$). As the results under exponential utility are independent of the policyholder's initial wealth $w_0$, we do not specify it further. When analyzing basis risk, we refer to
\begin{equation}\label{BasisRisk}
B_i := Y_i - S_i = e_{\g(\alpha^{*})}\bigl(S \big| \t = \theta_i\bigr) \mathds{1}_{\{\theta_i \in \T \}}  - S_i, \qquad i=1,\ldots,N , 
\end{equation}
instead of the asymmetric squared penalization underlying (\ref{DefBROpt}), to make positive and negative deviations distinguishable. 

\begin{figure}[t]
\center
\includegraphics[width = \textwidth, keepaspectratio]{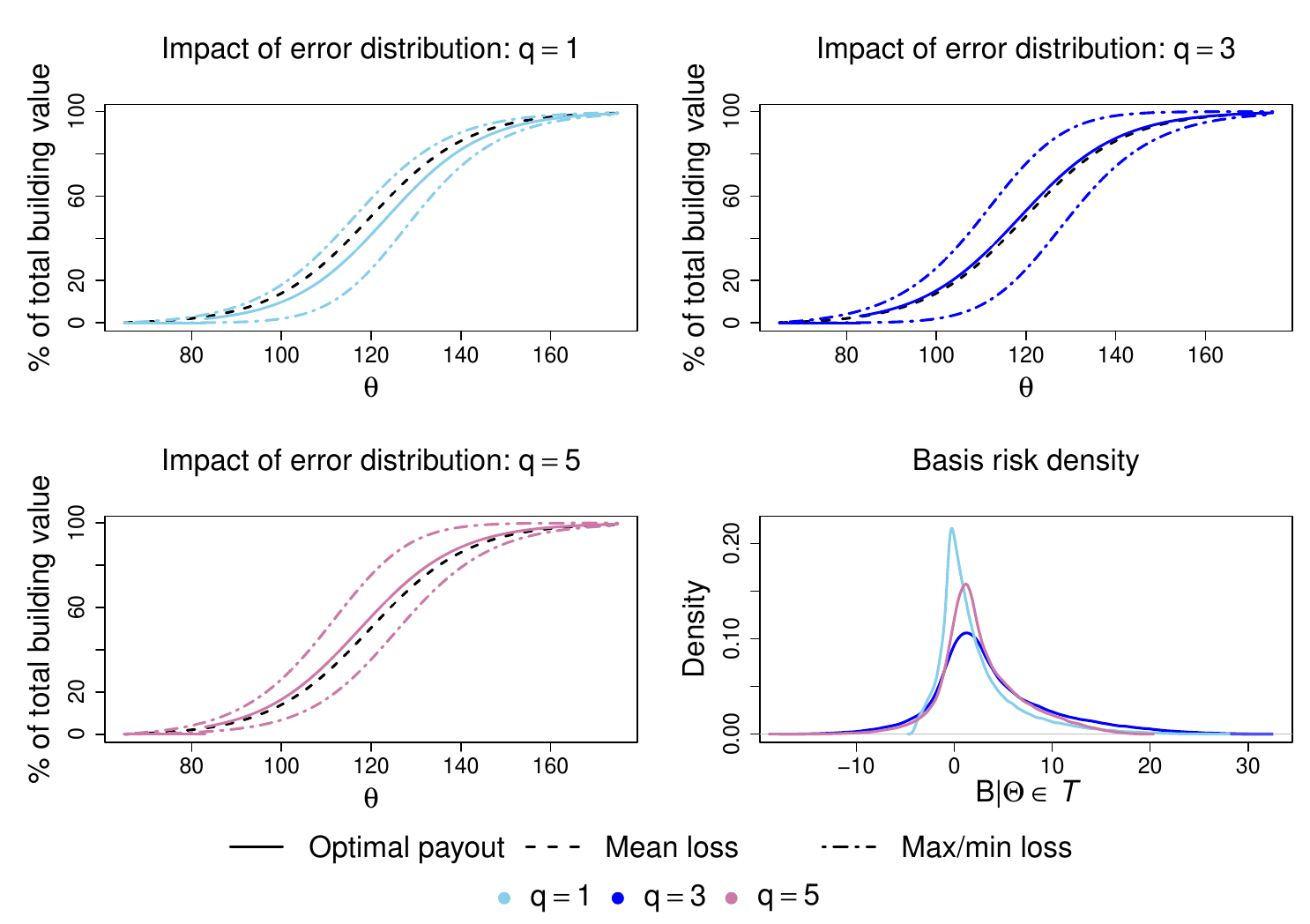}
\caption{Conditional mean, maximum, and minimum loss $S|\t$ versus optimal expectile-based payment scheme for noise parameters $p=3$, $q \in \{1,3,5\}$. Empirical densities of basis risk in the sense of \eqref{BasisRisk} are provided in the bottom right window.}
\label{VarError}
\end{figure}

We first consider the impact of the different error structures presented in Figure~\ref{Loss_Error_Distributions}. Algorithm~\ref{AlgoOptim} yields optimal basis risk weightings $(\a^{*,q = 1},\a^{*,q = 3},\a^{*,q = 5}) \approx (0.186, 0.587, 0.724)$. The resulting payment schemes clearly reflect the (a-)symmetry of the underlying errors and result in similar basis risk distributions, see Figure~\ref{VarError}.

\begin{figure}[t]
\center
\includegraphics[width = \textwidth, keepaspectratio]{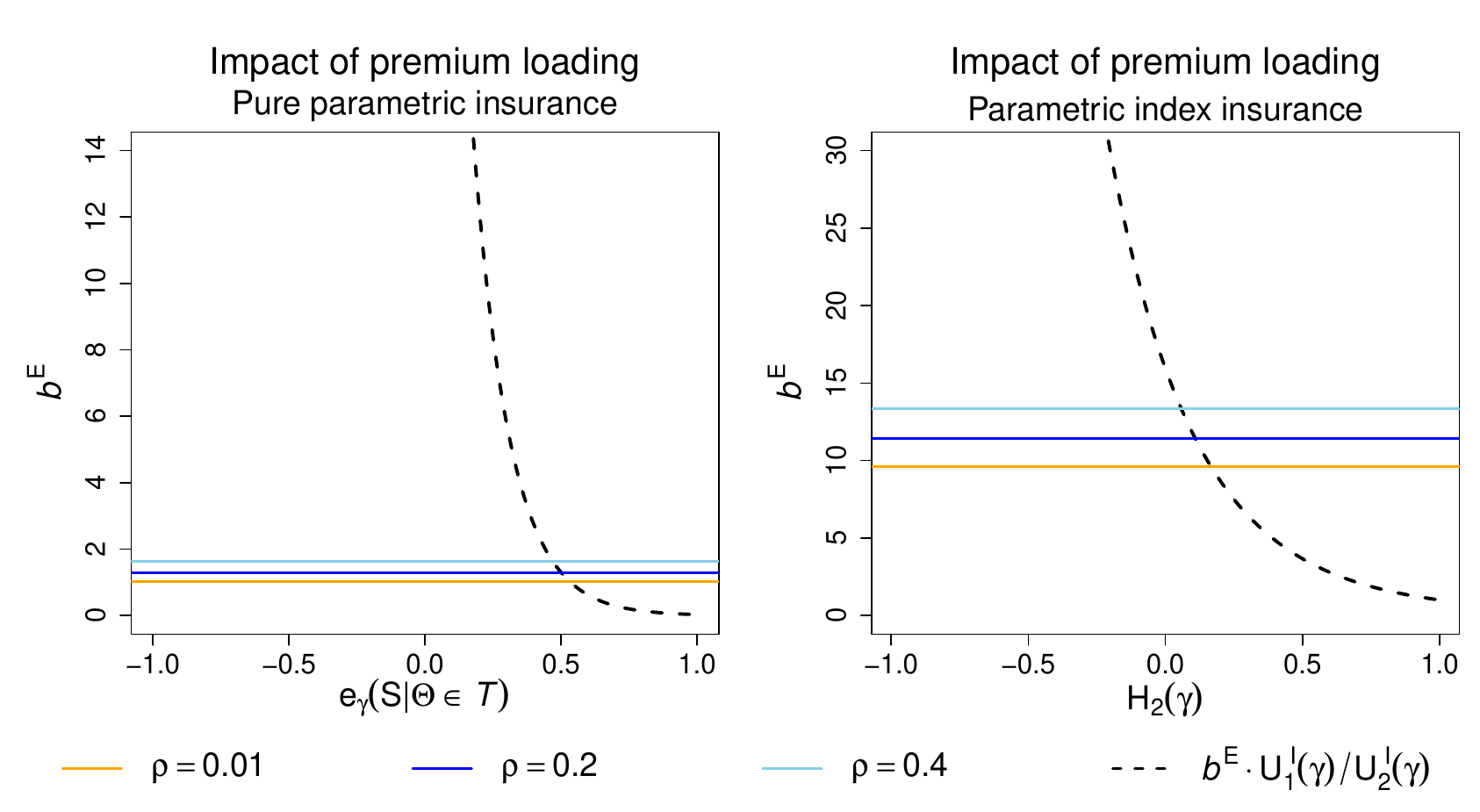}
\caption{Boundary conditions of Corollary~\ref{CorExp} (left) and Theorem~\ref{ExistencePI} (right) for premium loadings $\rho \in \{0.01,0.2,0.4\}$. Horizontal lines correspond to the right-, dotted lines to the left-hand side of (\ref{EU_UB}), resp. (\ref{UB_PI_EV}), evaluated at $k = e_\gamma(S|\t \in \T)$, resp. $k = H_2(\gamma)$. The optimal basis risk weighting $\a^*$ can be derived from the locations $e_{\gamma(\alpha^*)}(S|\t \in \T)$ and $k = H_2\bigl(\gamma(\alpha^*)\bigr)$ at which the lines intersect.} 
\label{VarRho}
\end{figure}

Next, we investigate the influence of the premium loading. Logically, we would \emph{ceteris paribus} expect an increase in $\rho$ to result in a lower $\a^*$, as higher premium costs should lead to a smaller optimal coverage. For pure parametric insurance, this behaviour is also supported by (\ref{OptEU}) through a monotonicity argument. Visualizing our simulation results in Figure~\ref{VarRho}, we can confirm the expected change in $\a^*$, corresponding to the shift of the intersection points of the dotted and colored lines at location $e_{\g(\a^*)}(S|\t \in \T)$, resp. $H_2\bigl(\g(a^*)\bigr)$, for both pure parametric and parametric index insurance.

\begin{figure}[t]
\center
\includegraphics[width = \textwidth, keepaspectratio]{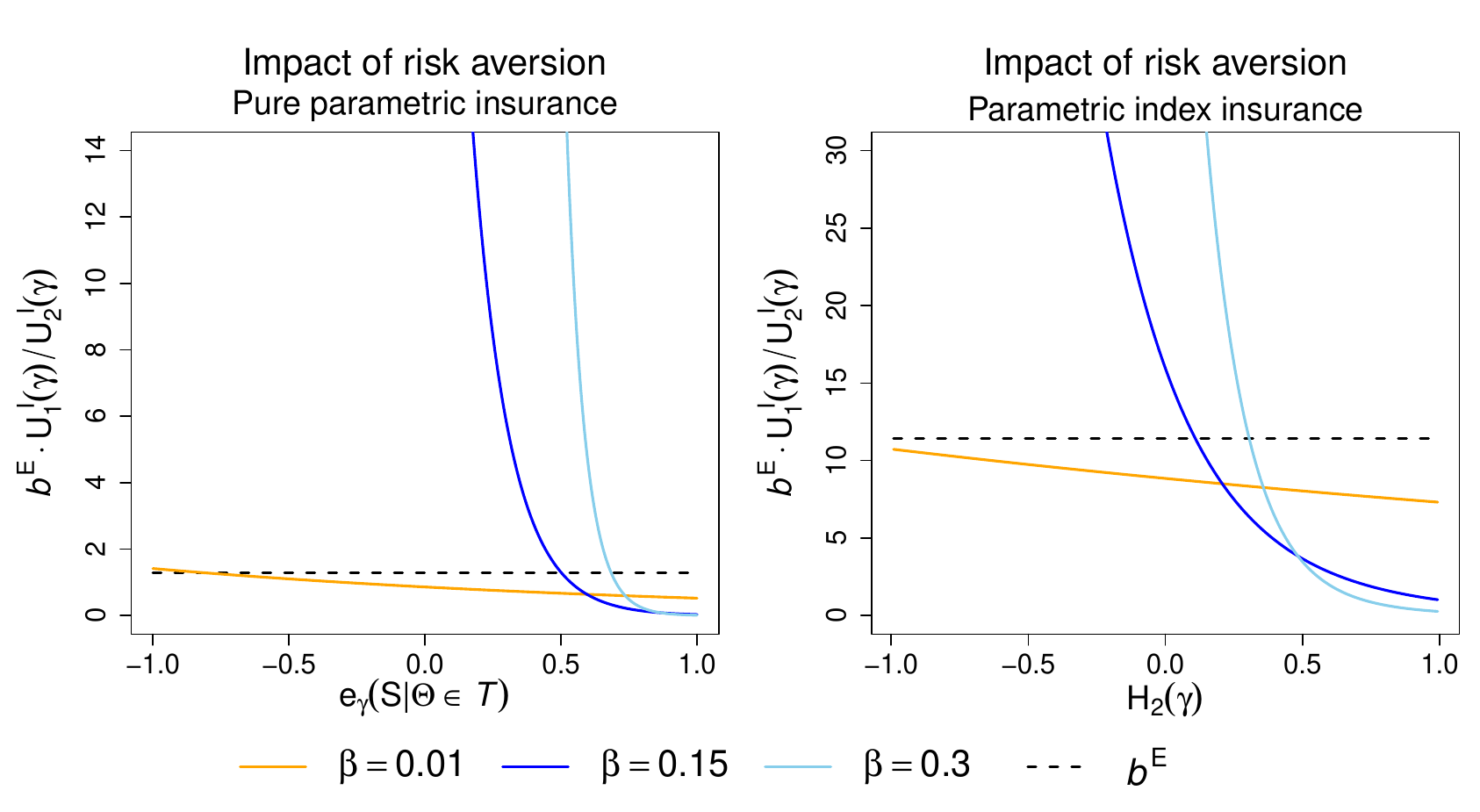}
\caption{Boundary conditions of Corollary~\ref{CorExp} (left) and Theorem~\ref{ExistencePI} (right) for risk aversion parameter $\beta \in \{0.01,0.15,0.3\}$. Horizontal lines correspond to the right-, colored lines to the left-hand side of (\ref{EU_UB}), resp. (\ref{UB_PI_EV}), evaluated at $k = e_\gamma(S|\t \in \T)$, resp. $k = H_2(\gamma)$. The optimal basis risk weighting $\a^*$ can be derived from the locations $e_{\gamma(\alpha^*)}(S|\t \in \T)$ and $H_2\bigl(\gamma(\alpha^*)\bigr)$ at which the lines intersect.} 
\label{VarBeta}
\end{figure}

Likewise, regarding the influence of the policyholder's risk aversion parameter $\beta$, it seems reasonable to expect more risk averse policyholders to opt for higher coverage and thus exhibit a larger optimal $\a^*$. As the impact of changing $\beta$ on the critical point condition is not immediate from our theoretical observations even for the simpler case of pure parametric insurance, we turn to the simulation results displayed in Figure~\ref{VarBeta}. For both types of parametric insurance under investigation, the hypothesized impact of $\beta$ on $\a^*$ can indeed be observed from the shift in intersection points. Further, the lack of an intersection point for $\beta = 0.01$ under parametric index insurance shows that the lower boundary condition (\ref{LB_PI_EV}) is violated.

In practice, payment schemes in parametric index insurance often take comparably simple forms like linear or step-functions \cite{Steinmann2023}.
In this spirit, we want to compare two piecewise linear payment schemes, whose functional form
\begin{equation}\label{PS_LinEx}
Y^\circ:= h^\circ(\t) = \min\bigl\{ \max\{0, \lambda^\circ(\t - 83) \}, v \bigr\}, \quad \circ \in \{\text{E},\text{U}\},
\end{equation}
is governed by the slope parameter $\lambda^\circ > 0$. The basis risk-optimal utility maximum $Y^\text{E}$ corresponds to the best approximation of the optimal expectile payment scheme $e_{\g(\a^*)}(S|\t)\Ind_{\{\t \in \T\}}$ for \mbox{$\a^*\approx 0.587$} as derived in our simulations. Following the comments in \cite{Maier2025}, this is achieved by numerically approximating
\begin{align*}
\lambda^\text{E} &:=\underset{\lambda>0}{\argmin} \; \E\bigl[ \g^* \bigl(S - Y^\text{E}\bigr)_+^2 + (1 - \g^*)^2 \bigl(S - Y^\text{E}\bigr)_-^2 \bigr] \\
&\; \approx \underset{\lambda>0}{\argmin} \; \frac{1}{n}  \sum_{i=1}^n \bigl[ \g^* \bigl(S_i - h^\text{E}(\theta_i)\bigr)_+^2 + (1 - \g^*)^2 \bigl(S_i - h^\text{E}(\theta_i)\bigr)_-^2 \bigr].
\end{align*}

\begin{figure}[t]
\center
\includegraphics[width = \textwidth, keepaspectratio]{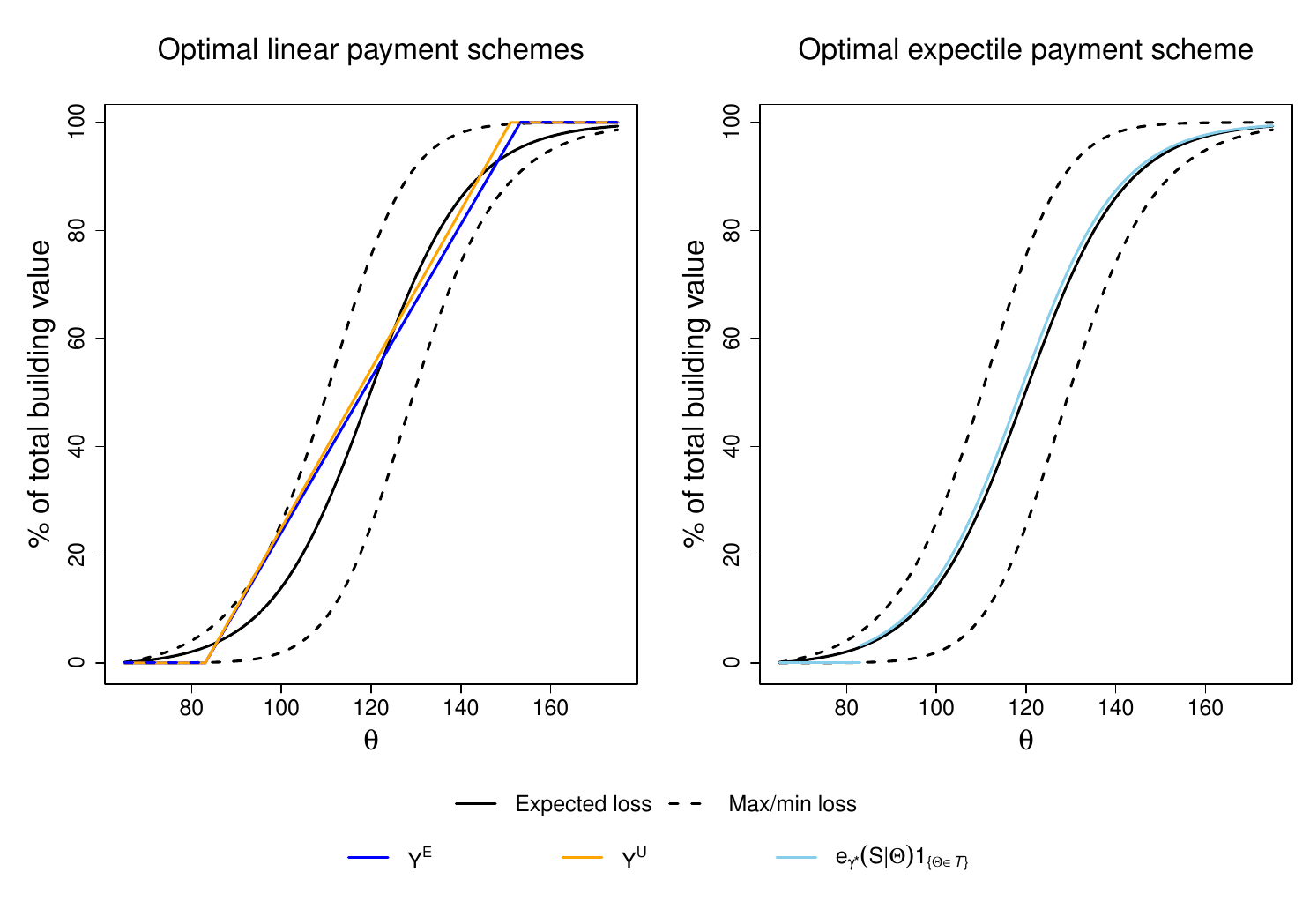}
\caption{Comparison of the mean, maximum, and minimum true conditional loss with piecewise linear payment schemes for basis-risk optimal utility maximization $Y^\text{E}$, resp. pure utility maximization $Y^\text{U}$, (left) and expectile-based payment scheme for parametric index insurance with optimal basis risk weighting $\a^* \approx 0.587$ (right).} 
\label{Lin_Payment}
\end{figure}

\pagebreak
For comparison, we consider the pure utility maximum $Y^\text{U}$ corresponding to
\begin{align*}
\lambda^\text{U} &:= \underset{\lambda>0}{\argmax}\; \E\bigl[u(w_0 - S + Y^\text{U} - (1 + \rho)\E[Y^\text{U}]) \bigr] \\
&\; \approx \underset{\lambda>0}{\argmax}  \; \frac{1}{n} \sum_{i=1}^n u\left(w_0 - S_i + h^\text{U}(\theta_i) - \frac{(1 + \rho)}{n} \sum_{j=1}^n h^\text{U}(\theta_j) \right).
\end{align*}
that does not (directly) take into account basis risk.\footnote{Especially, while negative basis risk still adversely impacts policyholder utility, positive basis risk is no longer considered detrimental (although some indirect penalization is still included through the premium).} Visualizing the resulting payment schemes in Figure~\ref{Lin_Payment}, we recognize that, while close, the lack of consideration for (positive) basis risk under $Y^\text{U}$ leads to more frequent and extensive overcompensation. Further, one can clearly recognize that the restriction to relatively simple payment schemes of form (\ref{PS_LinEx}) results in inferior coverage compared to the optimal expectile-based payment scheme, which is better able to mimic the true relationship between wind speed and losses. Especially, if one interprets $\mu(\t)$ as the ``average'' conditional loss around which the policyholder's true damages fluctuate due to individual randomness (represented by $Z_i$), the expectile-based payment scheme exposes the policyholder to far less \emph{design risk} (see \cite{Benami2021}) than $Y^\text{E}$ and $Y^\text{U}$.  

\subsection{Dependence aspects of parametric hurricane insurance}\label{SimDep}
Lastly, we extend the scope of our analysis to the topic of dependent losses, that - while not part of our derivations in Section~\ref{TheoOpt} - is indispensable from a real-world (portfolio) point of view. Thus, we consider the following four policyholders:
\begin{itemize}
\item Policyholder DW, as introduced in Section \ref{SimSetup}.
\item Policyholder BG, located at 28.04° N 82.42° W.
\item Policyholder FM, located at 26.64° N 81.87° W.
\item Policyholder FL, located at 26.12° N 80.13° W.
\end{itemize}

\begin{figure}[t]
\centering
\begin{minipage}[t]{0.45\textwidth}
\center
\includegraphics[width = 6cm, keepaspectratio]{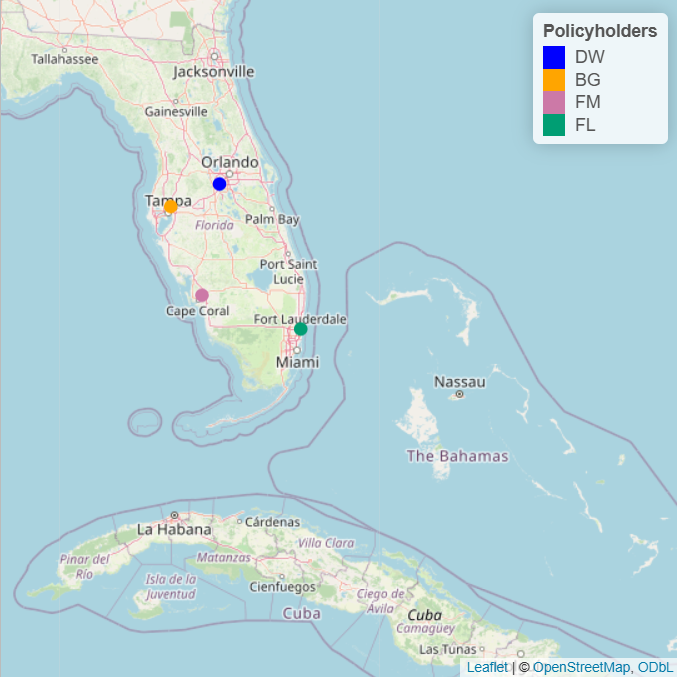}
\caption[Spatial distribution of the policyholders under investigation in Section~\ref{SimDep}. Map generated with R package \texttt{leaflet}.]{Spatial distribution of the policyholders under investigation in Section~\ref{SimDep}. Map generated with R package \texttt{leaflet}.\footnote{See footnote \ref{FNleaflet}.}} 
\label{Policyholder_Locations}
\end{minipage}
\hfill
\begin{minipage}[t]{0.45\textwidth}
\center
\includegraphics[width = 6cm, keepaspectratio]{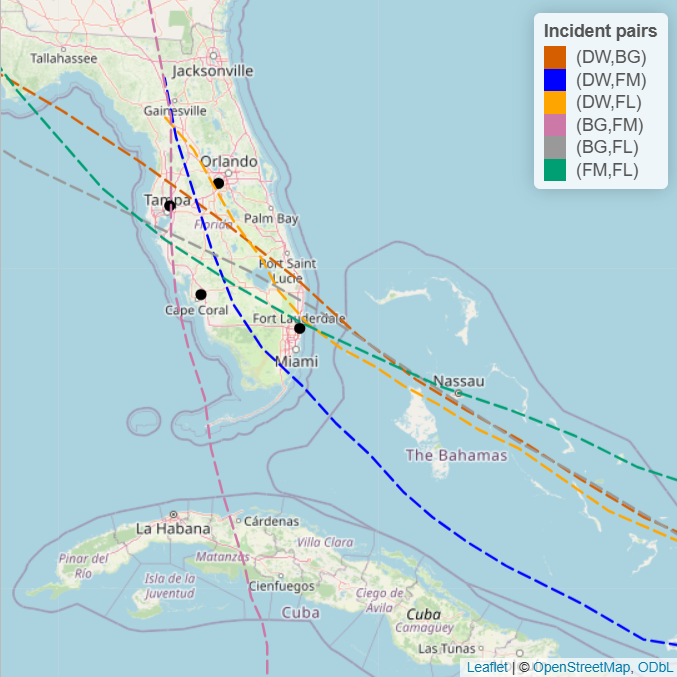}
\caption[Sample hurricane paths in STORM causing incidents at (at least) two policyholders. Map generated with R package \texttt{leaflet}.]{Sample hurricane paths causing incidents for (at least) two policyholders. Map generated with R package \texttt{leaflet}.\footnote{See footnote \ref{FNleaflet}.}} 
\label{Multi_Policyholder_Tracks}
\end{minipage}
\end{figure}

Figure~\ref{Policyholder_Locations} visualizes the policyholders' spatial distribution using the R package \texttt{leaflet}.\footnote{\label{FNleaflet}See \url{https://cran.r-project.org/package=leaflet} The package uses open data from OpenStreepMap, available under the Open Data Commons Open Database License (ODbL), see \url{https://www.openstreetmap.org/copyright/}.}. All policyholders are assumed as homogeneous w.r.t. the contract parameters $( r,\T, \rho, v) = (50\, \text{km}, [83,\infty), 0.2, 100)$ and possess the same risk aversion parameter $\beta = 0.15$. Following the steps outlined in Section~\ref{SimSetup}, we extract the joint maximum wind speeds $\bm{\theta}_i = (\theta_i^\text{DW},\theta_i^\text{BG},\theta_i^\text{FM},\theta_i^\text{FL})$ from all $n_\text{tot} = 107\,063$ tracks in the STORM data set, where $\theta_i^\circ := 0$ for policyholders whose trigger circle is not intersected by the $i$-th hurricane track. Policyholders' individual losses, i.e. the components of the joint loss vector $\bm{S}_i := (S_i^\text{DW},S_i^\text{BG},S_i^\text{FM},S_i^\text{FL})$, only depend on their respective maximum wind speed and follow (\ref{Model_Loss}). Especially, we assume independence of the error terms $Z_i^\circ \sim \text{Beta}(3,3)$ across policyholders, so that their spatial distribution is the only source of dependence between their losses.\footnote{One could, e.g., interpret $Z_i^\circ$ as describing random site- or even building-specific circumstances that cause deviations from the ``average'' S-shaped hazard function $\mu(\t_i^\circ)$.}

\pagebreak
We begin by analysing the (approximate) conditional incident / trigger probabilities
\begin{align*}
&p_{\text{inc}}^{\circ,*} := \P\left(\t^\circ > 0 \, \middle| \, \t^* > 0 \right), \qquad
&&p_{\text{trig}}^{\circ,*} := \P\left(\t^\circ \in \mathcal{T} \, \middle| \, \t^* \in \mathcal{T} \right),  \\
& \hat{p}_{\text{inc}}^{\circ,*} :=\sum_{i = 1}^{n_\text{tot}} \frac{\mathds{1}_{\{ \theta_i^\circ > 0 , \, \theta_i^* > 0 \}}}{\sum_{j = 1}^{n_\text{tot}} \mathds{1}_{\{ \theta_j^* > 0\}} }  , 
&&  \hat{p}_{\text{trig}}^{\circ,*} :=\sum_{i = 1}^{n_\text{tot}} \frac{\mathds{1}_{\{ \theta_i^\circ \in \mathcal{T} , \,\theta_i^* \in \mathcal{T} \}}}{\sum_{j = 1}^{n_\text{tot}} \mathds{1}_{\{ \theta_j^* \in \mathcal{T} \}} } ,
\end{align*}
displayed in Tables~\ref{Table_BayHit} and~\ref{Table_BayTrig}. The impact of the policyholders' spatial distribution can not only be observed through the difference in $\hat{p}_{\text{inc}}^{\circ,*}$ and $\hat{p}_{\text{trig}}^{\circ,*}$ between pairs, with (BG,FM), (FM,FL), and especially (DW,BG) exhibiting comparably larger values, but also through the asymmetry within pairs. While generally close for two given policyholders $(\circ,*)$, conditioning on the more southern one yields higher conditional incident and trigger probabilities. This could be attributed to the hurricanes' main direction, with tracks commonly crossing Florida from the south(-east), see Figure~\ref{Multi_Policyholder_Tracks} as well as \cite[Figure 1]{Bloemendaal2020}. Additionally, an interesting deviation from the ``common'' trend is observed for the pair (FM,FL): While conditional trigger probabilities are lower than conditional incident probabilities for all other pairs, we see a sizable increase of $\hat{p}^{\text{FM},\text{FL}}_\text{trig}$ and $\hat{p}^{\text{FL},\text{FM}}_\text{trig}$ compared to $\hat{p}^{\text{FM},\text{FL}}_\text{inc}$, resp. $\hat{p}^{\text{FL},\text{FM}}_\text{inc}$. Potential explanations for this could lie in a multitude of influencing factors used by \cite{Bloemendaal2020} in simulating maximum wind speeds and the differences in storm track movement required for joint incidents at FM and FL.\footnote{E.g., one would reasonably expect hurricanes that hit both locations to exhibit less change in latitude in the area of interest compared to those causing joint incidents for other policyholder pairs.} 

To investigate policyholders' bivariate (central) dependence, we make use of the well-known Kendall's $\tau$ as well as Chatterjee's $\xi$ \cite{Chatterjee2021}. The latter is able to also represent asymmetric dependencies (which seems fruitful in view of Tables~\ref{Table_BayHit} and~\ref{Table_BayTrig}) and is approximated by its (generalized) sample version\footnote{Due to the presence of ties in our data, we cannot make use of the simple formula \cite[Eq.\ (1)]{Chatterjee2021}.}

\begin{gather*}
\hat{\xi}^{\circ,*} := 1 - \frac{3\sum_{i=1}^{m-1} \left| r_{i+1} - r_i \right|}{2 \sum_{i = 1}^m l_i(m - l_i)} \\
r_i :=  \text{rank}\bigl(\theta^*_{(i)}\bigr) = \sum_{j=1}^{m} \mathds{1}_{\{\theta^*_{(i)} \ge \theta^*_{(j)} \}}, \qquad l_i := \sum_{j=1}^{m} \mathds{1}_{\{\theta^*_{(i)} \le \theta^*_{(j)} \}}, \nonumber
\end{gather*}
where $(\theta^\circ_{(1)},\theta^*_{(1)}),\ldots,(\theta^\circ_{(m)},\theta^*_{(m)})$ corresponds to ordering the paired observations $(\theta^\circ_i,\theta^*_i)_{i=1,\ldots,m}$ s.t. $\theta^\circ_{(1)} \le \theta^\circ_{(2)} \le \ldots \le \theta^\circ_{(m)}$. 

\begin{figure}[H]
\center
\includegraphics[width = \textwidth, keepaspectratio]{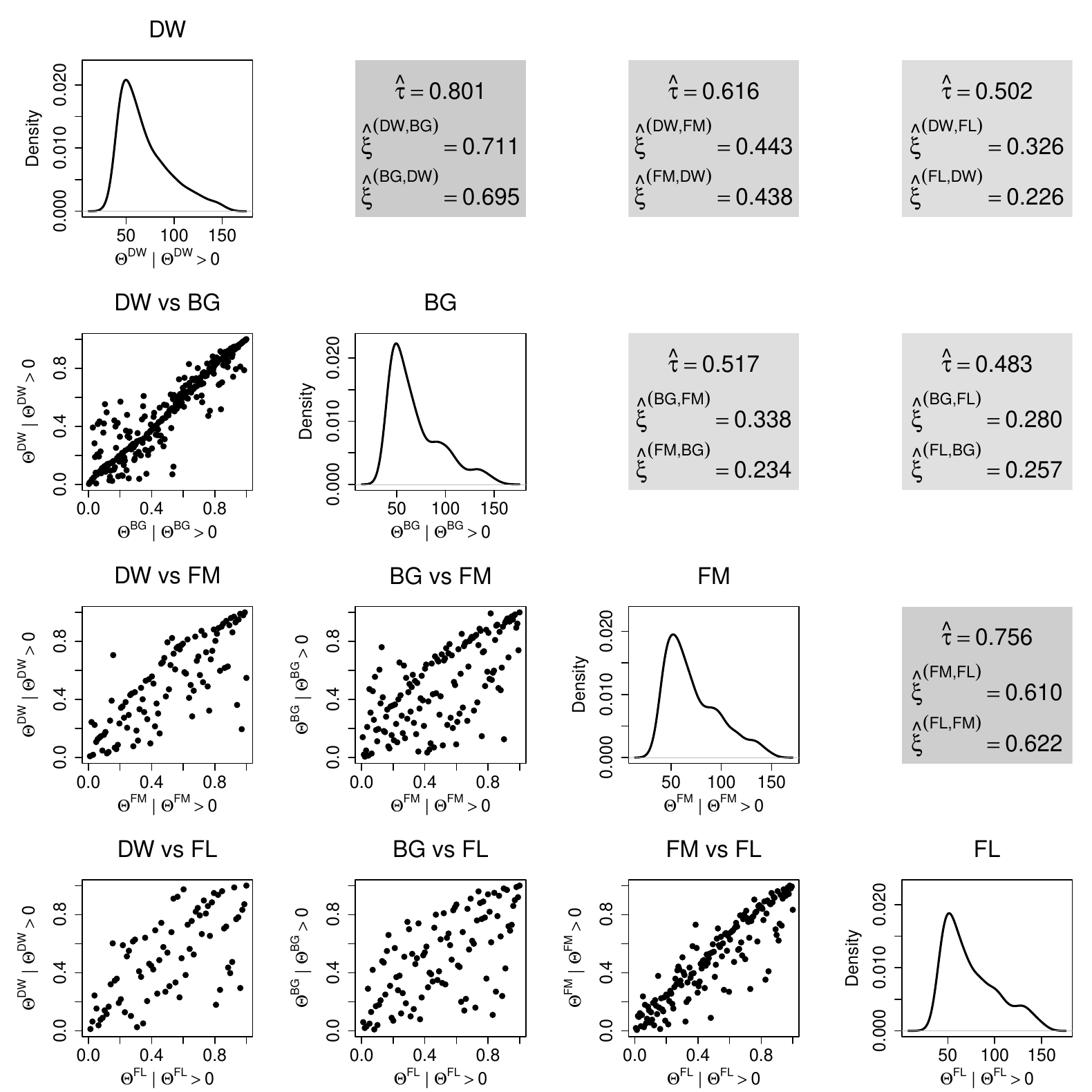}
\caption{(Bivariate) dependence analysis of wind speeds $(\t^\circ,\t^*)$ for policyholder pairs' joint incidents using rank transform scatterplots, Kendall's $\tau$, and Chatterjee's~$\xi$. Darker shades of grey correspond to higher values of $\hat{\tau}$. Univariate density plots for $(\t^\circ |\t^\circ > 0)$ are displayed on the diagonal.} 
\label{DependenceWind}
\end{figure}

Figure~\ref{DependenceWind} surveys rank-transform plots, $\hat{\tau}$, and $\hat{\xi}$ for all policyholder pairs, focusing on hurricanes resulting in joint incidents, i.e. $(\theta^\circ_j, \theta^*_j) \in (0,\infty)^2$. As is to be expected from hurricanes hitting both locations, maximum wind speeds for all policyholder pairs show clear signs of positive dependence. Additionally, policyholders with similar latitude, i.e. (DW,BG) and (FM,FL), exhibit stronger, rather symmetric dependencies. In contrast, pairs located on close longitudes, i.e. (BG,FM) and (DW,FL), have lower but considerably more asymmetric dependence. In the latter case, wind speeds recorded at the northern policyholder contain more information about the observations at the southern policyholder, that is normally hit earlier by the hurricane.\footnote{While reasonable, this is unfortunately rather undesirable from the point of view of a regulatory entity, who is decidedly more interested in predicting the hurricane's future path and impact. The same comments apply to traditional indemnity insurers who may want to issue preemptive warnings to policyholders along the expected hurricane path to reduce losses.}

For analysing the central dependence of basis risk, we first derive the optimal weighting $\a^*$ for the remaining policyholders analogously to the steps undertaken for DW in Section~\ref{SimSetup}.\footnote{Especially, the bootstrap sample of maximum wind speeds is generated solely based on hurricanes causing an incident at the policyholder of interest.} As policyholders have equal utilities and conditional loss distributions as well as comparably similar wind speed distributions, their optimal weightings
\begin{equation} \label{SimStudyWeights}
\left(\a^{*,\text{DW}}, \a^{*,\text{BG}}, \a^{*,\text{FM}},\a^{*,\text{FL}} \right) \approx (0.587, 0.578, 0.582, 0.577)
\end{equation}
are also close. 

\begin{table}[t]
\begin{center}
\begin{minipage}{0.45\linewidth}
\centering
\begin{tabular}{c|cccc}
\toprule
$\boldsymbol{\hat{p}_{\text{inc}}^{\circ,*}}$& \textbf{DW} & \textbf{BG} & \textbf{FM} & \textbf{FL} \\  
\midrule
\textbf{DW} & - & 0.383 & 0.137 & 0.105 \\
\textbf{BG} &0.375 &- & 0.186 & 0.131 \\
\textbf{FM} & 0.128 & 0.179 &- & 0.196 \\
\textbf{FL} & 0.095 & 0.122 & 0.190 &- \\
\bottomrule
\end{tabular}
\caption{Estimated conditional incident probabilites $\hat{p}_{\text{inc}}^{\circ,*}$ for the policyholders in Section~\ref{SimDep} (rounded to 3 digits).}\label{Table_BayHit}
\end{minipage}
\hfill
\begin{minipage}{0.45\linewidth}
\centering
\begin{tabular}{c|cccc}
\toprule
$\boldsymbol{\hat{p}_{\text{trig}}^{\circ,*}}$ & \textbf{DW} & \textbf{BG} & \textbf{FM} & \textbf{FL} \\
\midrule
\textbf{DW} & - & 0.367 & 0.092 & 0.082 \\
\textbf{BG} &0.350 &- & 0.155 & 0.112 \\
\textbf{FM} & 0.076 & 0.134 &- & 0.273 \\
\textbf{FL} & 0.059 & 0.084 & 0.238 &- \\
\bottomrule
\end{tabular}
\caption{Estimated conditional trigger probabilites $\hat{p}_{\text{trig}}^{\circ,*}$ for the policyholders in Section~\ref{SimDep} (rounded to 3 digits).}\label{Table_BayTrig}
\end{minipage}
\end{center}
\end{table}

\begin{figure}[H]
\center
\includegraphics[width = \textwidth, keepaspectratio]{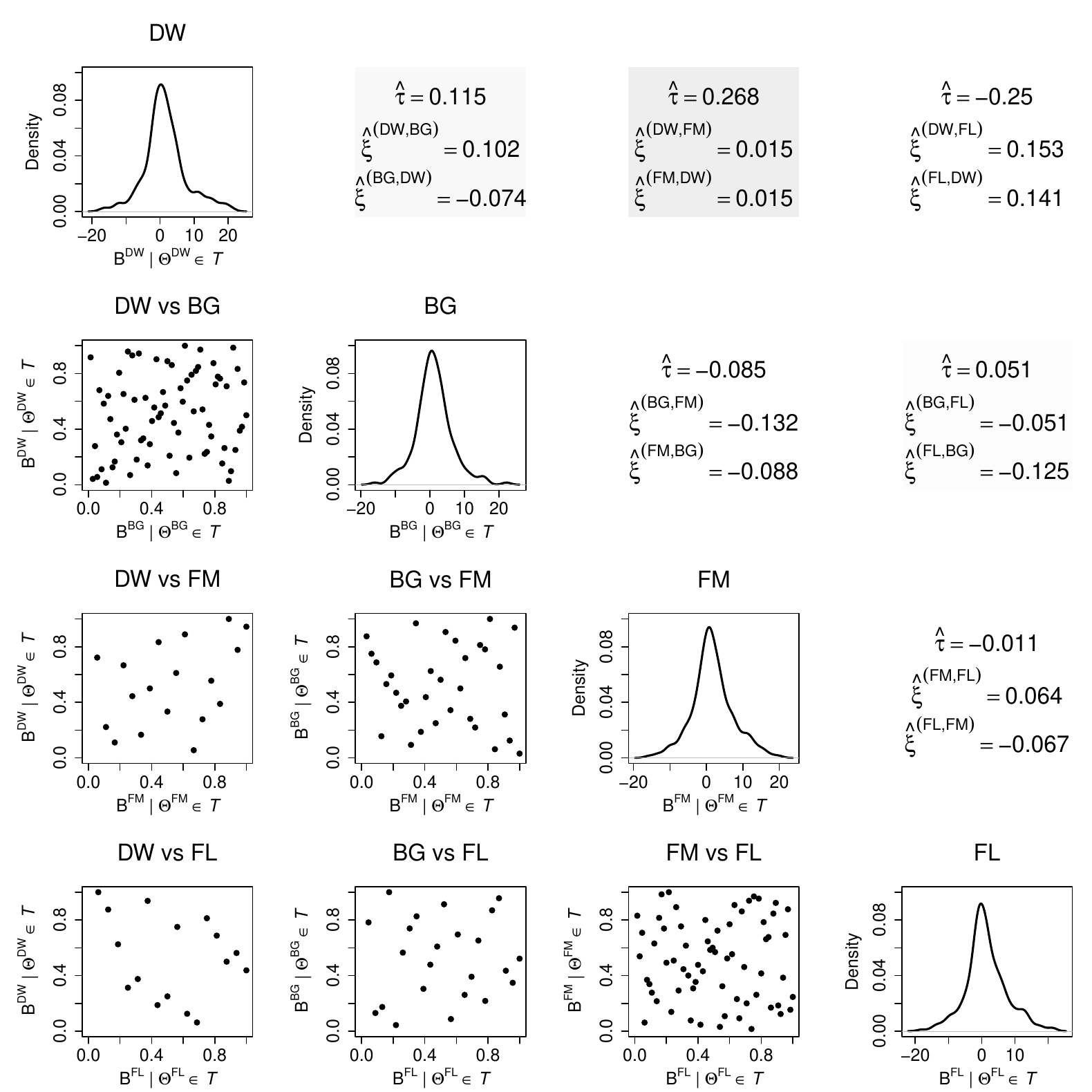}
\caption{(Bivariate) dependence analysis of basis risk $(B^\circ,B^*)$ as defined in (\ref{BasisRisk}) for hurricanes jointly triggering non-zero payments to policyholder pairs using rank transform scatterplots, Kendall's $\tau$, and Chatterjee's~$\xi$.  Darker shades of grey correspond to higher values of $\hat{\tau}$. Univariate density plots for $(B^\circ |\t^\circ \in \mathcal{T})$ are displayed on the diagonal. Note that estimated dependency measures may lack reliability for pairs with a low number of admissible observations.} 
\label{DependenceBR}
\end{figure}

Considering only hurricanes that trigger a non-zero payout to both policyholders in the pairs, i.e. $(\theta^\circ_i, \theta^*_i) \in \mathcal{T}^2$, Figure~\ref{DependenceBR} reports the rank-transform plots as well as $\hat{\tau}$ and $\hat{\xi}$.\footnote{While we show plots and estimates for all policyholder combinations, care should be taken when looking at pairs with a low number of data points, for whom $\hat{\tau}$ and $\hat{\xi}$ may lack reliability. We therefore base our comments mostly on the observations for (DW,BG) and (FM,FL), but our arguments should (by construction of the simulated observations) nonetheless be transferable to all other policyholder pairs.} In contrast to the strong correlation of wind speeds, while once again similar w.r.t. their marginal distributions, basis risk shows no signs of relevant dependence across policyholders. Recalling the simulation setup in the beginning of this section, this result is clearly in line with our expectations, as payments are close to the mean loss (recall (\ref{SimStudyWeights}) and $e_{0.5}(X) = \mathbb{E}[X]$) and basis risk is thus driven primarily by the \emph{independent} error terms $Z_i^\circ$. From the insurer's point of view, independence of basis risk across policyholders should be a desirable outcome, as joint (strong) negative experiences across policyholders (i.e. large negative basis risk realizations) may pose a substantial reputational risk that could endanger the parametric product's long-term viability.

To conclude our analysis, we investigate the tail dependence of policyholders' experienced wind speeds in an effort to better understand the accumulation risk associated with joint extreme hurricane events. We make use of the methodology described in \cite{Schmidt2006} to estimate the upper-tail dependence coefficient $\lambda^{\circ,*}_U$ for $(\t^\circ,\t^*)$ by 
\begin{align}
\hat{\lambda}^{\circ,*}_U &= \hat{\Lambda}_U^{\circ,*}(1,1) \, ,\label{UTDCest} \\
\hat{\Lambda}_U^{\circ,*}(x,y)& := \frac{1}{k} \sum_{j=1}^m \Ind_{\{ r_j^\circ > m - kx, \; r_j^* > m - ky \}}, \qquad 0 < k < m, \quad k,m \in \mathbb{N}\, , \label{EmpTailCop}
\end{align}
where $r_j^\circ$ and $r_j^*$ denote the rank of $\theta^\circ_j$ resp. $\theta^*_j$ in the observed sample $(\theta^\circ_j,\theta^*_j)_{j=1,\ldots,m}$. More details are given in Appendix~\ref{TDCMethodology}. 

Analogous to Figure~\ref{DependenceWind}, we consider only observations $(\theta^\circ_j,\theta^*_j) \subset(0,\infty)^2$. Table~\ref{ResultsTDC} summarizes our results and gives (approximate) 95\% confidence intervals for $\lambda_U$. Clearly, the strong dependence of wind speeds for joint incidents identified in our prior investigations is still present in the upper tails of the distributions. Especially, note that we can clearly reject tail independence ($\Leftrightarrow \lambda^{\circ,*}_U = 0$) for all policyholder pairs based on the (approximate) 95\% confidence intervals. Additional, while the point estimates $\hat{\lambda}^{\circ,*}$ show a similar trend in ordering to the dependence measures reported in Figure~\ref{DependenceWind}, no significant difference can be identified at the considered $5\%$ level. For completeness' sake, we performed the same estimation for $(B^\text{DW},B^\text{BG})$\footnote{For (FM,FL), the approximation procedure for the confidence interval was not applicable as the MLE for the Gumbel-Hougaard copula's parameter was close to 1, which causes the confidence interval to degenerate to a single point.} based on all observations $(\theta^\text{DW}_j,\theta^\text{BG}_j) \subset\T^2$. The resulting approximate 95\% confidence interval $[-0.085,0.285]$ supports the tail independence that should be expected from the i.i.d. assumption on the error terms $Z^\circ_j$.

To end this section on a positive note, we want to stress that the strong (tail) dependence of wind speeds (and thus of resulting payments under parametric insurance) was observed for joint incidents. However, the probability 
\begin{equation*}
\hat{p}^{\circ,*} := \frac{1}{n_\text{tot}} \sum_{i = 1}^{n_\text{tot}} \mathds{1}_{\{ \theta_i^\circ > 0,\, \theta_i^* > 0 \}} \approx \P(\t^\circ > 0, \t^* > 0)
\end{equation*}
for a hurricane to hit two of the comparably small trigger areas is quite low, with $\hat{p}^{\text{DW},\text{BG}} \approx 2.67 \cdot 10^{-3}$ being the maximum value observed among the pairs.\footnote{As the tracks in STORM correspond to $10\,000$ years (assumed i.i.d.) of hurricane activity, this translates to a joint incident for (DW,BG) occurring roughly every 35 years.} For non-zero payments to be triggered to both policyholders in a pair, the probabilities are even lower, with a maximum of approx. $6.73 \cdot 10^{-4}$. Therefore, while offering insurance against hurricanes requires careful consideration of (tail) dependence and resulting accumulation risk, especially when (unlike in this study) underwriting multiple contracts with overlapping / shared trigger area, it still presents a viable area for (parametric) insurance solutions. 

\begin{table}[t]
\begin{center}
\begin{tabular}{c|cc|c|c }
\toprule
$\boldsymbol{(\circ,*)}$ & $\boldsymbol{m}$ & $\boldsymbol{k}$ & $\boldsymbol{\hat{\lambda}^{\circ,*}_U}$ & \textbf{95\% confidence interval}  \\
\midrule
(DW,BG) & 286 & 40 & 0.850 & $[0.766,0.934]$   \\
(DW,FM) & 102 & 35 & 0.686 & $[0.559,0.813]$  \\
(DW,FL) & 78 & 33 & 0.697 & $[0.552,0.842]$  \\
(BG,FM) & 142 & 50 & 0.700 & $[0.586,0.814]$ \\
(BG,FL) & 100 & 35 & 0.657 & $[0.517,0.797]$ \\
(FM,FL) & 156 & 26 & 0.731 & $[0.607,0.855]$ \\ 
\bottomrule
\end{tabular}
\caption{Non-parametric estimates for the bivariate upper-tail dependence coefficient (\ref{UTDCest}), based on parameters $m$, $k$, and approximate 95\% confidence intervals derived from (\ref{NormalApproxTDC}) for wind speeds of joint incidents at policyholder pairs.} \label{ResultsTDC}
\end{center}
\end{table}

\section{Conclusion}\label{Conclusion}

Appropriate design of the payment schemes that govern compensation in parametric insurance based on an underlying index plays a vital role for the long term viability of these risk transfer solutions. While conditional expectiles of policyholders' true losses have been shown by \cite{Maier2025} to be well-suited for this task, their level (and thus the overall indemnification) strongly depends on the importance $\a$ of negative basis risk relative to positive basis risk. As this weighting may not be known a priori, implementation of expectile payment schemes requires a way to determine it appropriately, which is where our main contribution lies.

As a rational policyholder's basis risk weighting should conform to the ultimate goal of utility-maximization, we can identify $\a$ by considering the behaviour of the policyholder's expected utility. Using differentiability of expectiles, we derive that existence of the optimal $\a$ under a given premium principle is characterized by two boundary conditions. Further, if it exists, the optimal basis risk weighting is always unique. In addition to supporting the implementation of expectile payment schemes, numerical calculation of the optimal $\a$ allows investigation of (hidden) influencing factors, like risk aversion and premium. If no parametric contract is optimal, we provide sufficient conditions for the policyholder to strictly prefer no insurance or full indemnity insurance over all index insurance offerings. 

Our results cover the expected value, variance, and standard deviation premium principle in pure parametric insurance as well as the expected value and variance premium principle in parametric index insurance. For the latter, a lack of general (dominance) results for expectiles' second order derivatives requires an assumption in the form of separable first order derivatives. If conditional losses have uniformly bounded variance, one can equivalently assume that conditional loss distributions belong to a location-scale family. This encapsulates both commonly used claim size distributions and familiar regression approaches. To the best of our knowledge, we are the first to note this characterization of separable expectile derivatives.

A simulation study on parametric cat-in-a-circle hurricane insurance, a popular area of application for index insurance, visualizes our insights and allows for investigating the influence of policyholder and contract characteristics. We observe $\a^*$ to be increasing in policyholder's risk aversion and decreasing in the contract's premium loading, a rather intuitive behaviour. Additionally, comparison with pure utility-maximization in the context of piecewise linear payment schemes shows that the direct penalization of positive basis risk inherent to the expectile-based approach incentivizes lower payments to reduce overcompensation. When considering a portfolio of policyholders, index realizations exhibit strong central and tail dependency, but correlation of basis risk can be kept low if payment schemes have small design risk.

Anticipating future research, there are many promising opportunities for extending our results on the optimal design of parametric insurance. As noted in \cite{Maier2025}, expectile-based parametric insurance does not consider the question of optimal index design and selection, motivating efforts to combine our approach with insights from this area of research (see, e.g., \cite{Stigler2023}). Additionally, while investigating the optimal $\a$ on an individual level is appropriate for primary insurance contracts, alternative uses of index insurance (e.g., parametric risk sharing among nation actors\footnote{For some examples in the context of natural catastrophes, we refer to \cite[Section 2.4.4]{Jerry2023}.}) may require taking a joint perspective. Lastly, investigating the behaviour of higher-order expectile derivatives could provide valuable insights and help extend our results to distributions outside the location-scale setting.

\vspace{0.5cm}
\noindent \textbf{Acknowledgments.} We thank An Chen for her helpful comments in preparing this manuscript. Additionally, Markus Johannes Maier gratefully acknowledges financial support by Allianz SE.

\vspace{0.3cm}
\noindent \textbf{Data availability statement.} The STORM data set used in Section~\ref{SimStudy}'s simulation study is available for download at \url{https://doi.org/10.4121/12706085.v4}, see \cite{BloemendaalData}.

\newpage
\vspace*{1cm}

 \newpage
\vspace*{1cm}
\newpage


\newpage
\begin{appendix}

\section{Additional proofs and derivations}

\subsection{Monotonicity of $\mathcal{V}_1$ and $\mathcal{V}_2$ in pure parametric insurance with variance premium principle}\label{PPVarDyn}
The behaviour of $\mathcal{V}_1$ and $\mathcal{V}_2$ becomes apparent by explicitly calculating 
\begin{align*}
\mathcal{V}'_1(\g) &= \underbrace{e'_{\g}(S|\t \in \T) \P(\t \in \T) \bigl(1 - r(\g) \bigr)^2}_{\ge 0} \, \underbrace{\E\left[u''\left(w_0 - S +  \bigl(1 - R(\g) \bigr) e_{\g}(S|\t \in \T) \right)| \t \in \T \right]}_{< 0} \\
& \qquad - \underbrace{\P(\t \in \T) \, r'(\g)}_{> 0} \underbrace{\E\left[u'\left(w_0 - S +  \bigl(1 - R(\g) \bigr) e_{\g}(S|\t \in \T) \right)| \t \in \T \right]}_{> 0} < 0,\\
\mathcal{V}'_2(\g) &=  - \underbrace{e'_{\g}(S|\t \in \T) \, \P(\t \notin \T) \, r^2(\g)}_{>0}  \underbrace{\E\left[u''\bigl(w_0 - S - R(\g) e_{\g}(S|\t \in \T)\bigr) | \t \notin \T \right]}_{<0} \\
& \qquad + \underbrace{\P(\t \notin \T) \, r'(\g)}_{> 0}  \underbrace{\E\left[u'\bigl(w_0 - S - R(\g) e_{\g}(S|\t \in \T)\bigr) | \t \notin \T \right]}_{> 0} > 0,\\
r'(\g)&=2\rho \P(\t \in \T)  \P(\t \notin \T) e'_\g(S|\t \in \T)>0.
\end{align*}

\subsection{Optimal insurance decision under violated boundary conditions}

\subsubsection{Remark \ref{PPVioBound}: Pure parametric insurance}\label{PI_VioB_proof}\label{PP_VioB_proof}

We begin with examining the case of violated lower boundary condition (\ref{LB_PP_ESD}), resp. (\ref{LB_PP_Var}). As $\essinf(S|\t \in \T) = 0$ implies $\lim_{\g \searrow 0} \U(\g) = \U_0$ (see Theorem~\ref{PropExp}) by continuity of $\U(\cdot)$, we only need to consider $\essinf(S|\t \in \T) > 0$. We can relate this problem back to the continuity argument from before by defining
\begin{equation*}
\tilde{\U}(x):= \E[u(w_0 - S + x\mathds{1}_{\{\t \in \T\}} - \pi_x)],
\end{equation*}
where $\pi_x$ is the premium corresponding to the payment $x\mathds{1}_{\{\t \in \T\}}$ under the principles (\ref{EV_Premium}) - (\ref{Var_Premium}). Clearly, as $e_\g(S|\t \in \T)$ is strictly increasing and continuous, for any  $\g \in (0,1)$ exists exactly one $x \in \bigl( \essinf(S|\t \in \T) , \, \esssup(S|\t \in \T) \bigr)$ s.t. $\tilde{\U}(x) = \U(\g)$ and vice versa. Additionally, the behaviour of $\tilde{U}(x)$ can be derived analogously to the steps undertaken in the proof of Theorem~\ref{ExistencePP}. Therefore, boundary conditions (\ref{LB_noInsurance_EVSD}) and (\ref{LB_noInsurance_Var}) imply that $\tilde{\U}(x)$ is strictly decreasing on $(0,\infty)$. As $\tilde{\U}(0)$ is equivalent to opting for no insurance, this concludes our considerations.

For comparison to full coverage indemnity insurance in case of violated upper boundary condition (\ref{UB_PP_ESD}), resp. (\ref{UB_PP_Var}), we note that strict concavity of $u$ and Jensen's Inequality let us write
\begin{equation*}
\U(\g) = \E[u(w_0 - S + Y^*_\g - \pi_\g )] < u\bigl(w_0 - \E[S] + \E[Y^*_\g] - \pi_\g  \bigr).
\end{equation*}
We can now compare the above to the \emph{deterministic} utility
\begin{equation*}
\E[u(w_0 - S + S - \pi_\text{I})] = u (w_0 - \pi_\text{I}) =: \U_\text{I},
\end{equation*}
resulting from a full coverage indemnity contract. Recalling the payment scheme (\ref{PS_PP}) as well as the premium principles (\ref{EV_Premium}) - (\ref{Var_Premium}), increasingness of $u$ yields:
\begin{itemize}
\item Expected value principle:
\begin{align*}
&e_\g(S|\t \in \T) > \frac{\rho_\text{I}}{\rho} \frac{\E[S]}{\P(\t\in\T)} \\
& \quad \Leftrightarrow \rho\, \E\bigl[e_\g(S|\t \in \T) \mathds{1}_{\{\t \in \T\}}\bigr] >  \rho_\text{I}\, \E[S]\\
& \quad \Leftrightarrow  w_0 - \E[S] + \E[Y^*_\g] - (1+\rho)\E[Y^*_\g] < w_0 - (1 + \rho_\text{I})\E[S]\; \Rightarrow \; \U(\g) < \U_\text{I}\, .
\end{align*}
\item Standard deviation principle:
\begin{align*}
&e_\g(S|\t \in \T) > \left(\frac{\rho_\text{I}}{\rho}\right)^2  \frac{\Var(S)}{\Var\left(\mathds{1}_{\{\t \in \T\}} \right)} \\
& \quad \Leftrightarrow \rho\sqrt{ \Var \left(e_\g(S|\t \in \T) \mathds{1}_{\{\t \in \T\}}\right)} >  \rho_\text{I} \sqrt{\Var(S)}\\
& \quad \Leftrightarrow  w_0 - \E[S] + \E[Y^*_\g] - \left( \E[Y^*_\g] + \rho \sqrt{\Var(Y^*_\g) } \right) < w_0 - \left( \E[S] + \rho_\text{I} \sqrt{\Var(S)} \right)\\
&\quad \Rightarrow \U(\g) < \U_\text{I}\, .
\end{align*}
\item Variance principle:
\begin{align*}
&\bigl(e_\g(S|\t \in \T)\bigr)^2 > \frac{\rho_\text{I}}{\rho}  \frac{\Var(S)}{\Var\left(\mathds{1}_{\{\t \in \T\}} \right)} \\
& \quad \Leftrightarrow \rho \Var \left(e_\g(S|\t \in \T) \mathds{1}_{\{\t \in \T\}}\right) >  \rho_\text{I} \Var(S)\\
& \quad \Leftrightarrow  w_0 - \E[S] + \E[Y^*_\g] - \Bigl( \E[Y^*_\g] + \rho \Var(Y^*_\g)  \Bigr) < w_0 - \Bigl( \E[S] + \rho_\text{I} \Var(S) \Bigr)\\
&\quad \Rightarrow \U(\g) < \U_\text{I} \, .
\end{align*}
\end{itemize}
As $\U(\g)$ is increasing on $(0,1)$, the criteria of Remark \ref{PPVioBound} follow from $\g \nearrow 1$ and Theorem~\ref{PropExp}, Property~\ref{LimitExp}.

\subsubsection{Remark \ref{PIVioBound}: Parametric index insurance}\label{PI_VioB_proof}

If $\essinf(S|\t= \theta) = 0$ for almost all $\theta \in \T$, Property~\ref{LimitExp} of Theorem~\ref{PropExp} implies that $\U(\g)$ converges to the utility of no insurance, $\U_0 := \E[u(w_0 - S)]$. Thus, the first statement of Remark~\ref{PIVioBound} follows directly from the strict decreasingness of $\U(\g)$ on $(0,1)$ when (\ref{LB_PI_EV}) resp. (\ref{UB_PI_V}) are violated. 

For comparison to full coverage indemnity insurance in case of violated upper boundary condition (\ref{UB_PI_EV}) resp. (\ref{UB_PI_V}), we mirror our approach from Appendix \ref{PP_VioB_proof}. Jensen's Inequality and increasingness of $u$ yield:
\begin{itemize}
\item Expected value principle:
\begin{align*}
&\E\left[e_\g(S|\t)\middle| \t \in \T\right] > \frac{\rho_\text{I}}{\rho \, \P(\t\in\T)} \E[S] \\
& \quad \Leftrightarrow \rho\, \E\left[e_\g(S|\t) \mathds{1}_{\{\t \in \T\}}\right] > \rho_\text{I} \, \E[S]\\
& \quad \Leftrightarrow  w_0 - \E[S] + \E[Y^*_\g] - (1+\rho)\E[Y^*_\g] < w_0 - (1 + \rho_\text{I})\E[S]\; \Rightarrow \; \U(\g) < \U_\text{I}\, .
\end{align*}
\item Variance principle:
\begin{align*}
&\Var \left(e_\g(S|\t) \mathds{1}_{\{\t \in \T\}}\right) > \frac{\rho_\text{I}}{\rho} \Var(S) \\
& \quad \Leftrightarrow \rho \Var \left(e_\g(S|\t) \mathds{1}_{\{\t \in \T\}}\right) > \rho_\text{I} \Var(S)\\
& \quad \Leftrightarrow  w_0 - \E[S] + \E[Y^*_\g] - \Bigl( \E[Y^*_\g] + \rho \Var(Y^*_\g)  \Bigr) < w_0 - \Bigl( \E[S] + \rho_\text{I} \Var(S) \Bigr) \\
&\quad \Rightarrow \U(\g) < \U_\text{I} \, .
\end{align*}
\end{itemize}
Here, $\U_\text{I} := u (w_0 - \pi_\text{I})$ again denotes the utility of full coverage indemnity insurance. As $\U(\g)$ is increasing on $(0,1)$, the criteria of Remark \ref{PPVioBound} follow from $\g \nearrow 1$ and Theorem~\ref{PropExp}, Property~\ref{LimitExp}.

\subsection{Proof of Proposition~\ref{SepDerivProp}}\label{ProofExpDeriv}
Proof of (\ref{proof1}):\\
``$\Leftarrow$'': Trivial.\\
``$\Rightarrow$'': Note that $e_{0.5}(S|\t)=\E[S|\t]$. Now, for any $\g \in (0,1)$:
\begin{equation*}
e_\g(S|\t) = \int_{0.5}^{\g} e'_{\tilde{\g}}(S|\t)\, d\tilde{\g} + \E[S|\t] = h_1(\t)\underbrace{\int_{0.5}^\g h_2(\tilde{\g}) \, d\tilde{\g}}_{=:\, H_2(\g)} + \underbrace{\E[S|\t]}_{=:\, H_3(\t)}.
\end{equation*}
\\
Proof of (\ref{proof2}):\\
``$\Leftarrow$'': Follows directly from Property \ref{CondCI} in Lemma~\ref{PropCondExp}. Especially, this implies $h_1(\t) \in \L^\infty(\t)$.\\
``$\Rightarrow$'': Define for a fixed $\theta \in \t(\Omega)$ 
\begin{align*}
Z &:=_d \left(S|\t = \theta \right) \in \L^2, \\
\mu(\t) &:= H_3(\t) - \frac{h_1(\t)H_3(\theta)}{h_1(\theta)} \in \L^2(\t),\\
\sigma(\t) &:= \frac{h_1(\t)}{h_1(\theta)} \in \L^\infty(\t).
\end{align*}
Then, $Z \perp \t$ holds and Property \ref{CondCI} in Lemma~\ref{PropCondExp} yields
\begin{align*}
e_\g\bigl(\mu(\t) + \sigma(\t)\,Z\big|\t\bigr) &= \mu(\t) + \sigma(\t)\,e_\g(Z|\t)\\
&=  H_3(\t) - \frac{h_1(\t)H_3(\theta)}{h_1(\theta)} + \frac{h_1(\t)}{h_1(\theta)} \bigl(h_1(\theta)H_2(\g) + H_3(\theta) \bigr)\\
&= h_1(\t)H_2(\g)+H_3(\t) = e_\g(S|\t).
\end{align*}
Since the distribution of a rv is characterized by its expectiles, see (\ref{ExpDerivative}), we can conclude \mbox{$S|\t=_{d} \mu(\t) + \sigma(\t)\,Z$}.

\section{Optimal $\a^*$ under violation of Assumption~\ref{SepDerivAssum}}\label{PINoAssum}
\begin{figure}[h]
\center
\includegraphics[width = \textwidth, keepaspectratio]{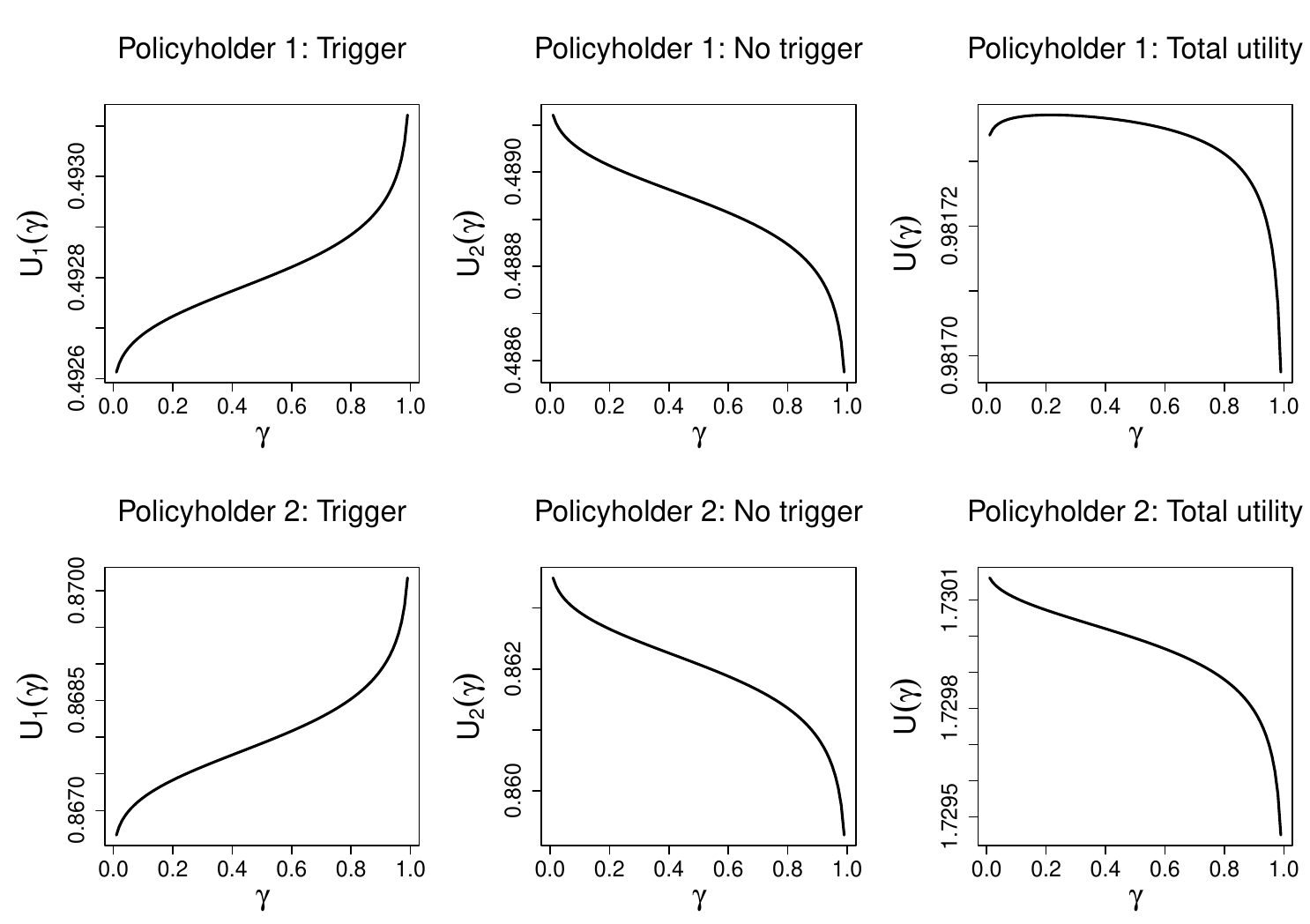}
\caption{Expected (sub-)utilities $\U_1(\g)$, $\U_2(\g)$, and $\U(\g)$ for policyholders $k=1,2$ under violation of Assumption~\ref{SepDerivAssum} based on $10^5$ simulation runs. A utility-optimal basis risk weighting $\a^*$ exists if and only if $\U(\g)$ admits a maximum on $(0,1)$.}
\label{CompareUtility}
\end{figure}

We investigate the expected utility of expectile-based parametric index insurance for policyholders $k=1,2$ in the following setting:
\begin{gather*}
\t \sim \text{Uniform}(2,4), \qquad (S^{(k)}|\t = \theta) \sim \text{Gamma}\bigl( r(\theta), \theta \bigr), \qquad r(\theta) = 3\mathds{1}_{\{\theta \le 3.5\}} + 3.5\mathds{1}_{\{\theta > 3.5\}}, \\
\T = (3,4], \qquad \pi^{(k)}_\g = \left(1+\rho^{(k)}\right)\mathbb{E}[Y_\g], \qquad u^{(k)}(x) = \frac{x^{1 - \eta^{(k)}} - 1}{1 - \eta^{(k)}},\\
\rho^{(1)} = 0.05, \rho^{(2)} = 0.1, \qquad  \eta^{(1)} = 2, \eta^{(2)} = 1.5, \qquad w_0^{(1)} = w_0^{(2)} = 65.
\end{gather*}
Clearly, the regime change at $\theta = 3.5$ violates Assumption~\ref{SepDerivAssum}. Nonetheless, policyholder 1 still exhibits a convex expected utility $\U$ (see Figure~\ref{CompareUtility}) and thus admits an optimal $\a^*$ in the sense of (\ref{EqOpt}). In contrast, while we observe the same monotonicity in $\U_1$ and $\U_2$ for policyholder 2, no optimal basis risk weighting exists in this case.

\section{Estimation of $\lambda_U^{\circ,*}$: Methodology and approximations}\label{TDCMethodology}

For the reader's convenience, we shortly summarize the key aspects of Schmidt and Stadtm\"uller \cite{Schmidt2006}'s methodology used in Section~\ref{SimDep}. The non-parametric estimator (\ref{UTDCest}) relies on approximating the \emph{upper tail copula} 

\begin{equation*}
\Lambda_U(x,y) = \lim_{t\to \infty} t \,  \bar{C}\left(\frac{x}{t},\frac{y}{t} \right) = x + y + \lim_{t\to \infty} t \left[ C\left(1 - \frac{x}{t},1 - \frac{y}{t} \right) - 1\right],
\end{equation*}
where ($\bar{C}$) $C$ is the (survival) copula of the random vector $(\t^\circ,\t^*)$, by (\ref{EmpTailCop}). The quality of this approximation depends on the parameter $k \in \{1,\ldots,m\}$, which is chosen by identifying a plateau in the graph of $k \mapsto \hat{\lambda}_U^{\circ,*}$ after appropriate kernel smoothing.

For $k,m$ large enough and under certain regularity conditions, \cite[Corollary 5]{Schmidt2006} shows asymptotic normality of $\hat{\lambda}_U^{\circ,*}$, i.e.
\begin{align}
&\sqrt{k} \bigl( \hat{\lambda}_U^{\circ,*} - \lambda_U^{\circ,*} \bigr) \approx_d \mathcal{N}(0,\sigma_U^2), \label{NormalApproxTDC} \\
&\sigma_U^2 = \Lambda_U^{\circ,*}(1,1) + \left(\frac{\partial}{\partial x} \Lambda_U^{\circ,*}(x,y) \right)^2 + \left(\frac{\partial}{\partial y} \Lambda_U^{\circ,*}(x,y) \right)^2 \label{sigU} \\
&\qquad \qquad+ 2\Lambda_U^{\circ,*}(1,1) \left[\left(\frac{\partial}{\partial x} \Lambda_U^{\circ,*}(x,y) - 1\right)\left(\frac{\partial}{\partial y} \Lambda_U^{\circ,*}(x,y) - 1\right) -1\right]. \nonumber
\end{align}
As the partial derivatives of $\Lambda_U^{\circ,*}(x,y)$ are unknown and their estimation is made difficult by the small number of observations in the tails, it is suggested to approximate $\sigma_U^2$. To that end, one chooses \emph{``a simple but flexible parametric copula''} \cite[p. 324]{Schmidt2006} for which one can calculate $\sigma_U^2(\eta)$ and replaces $\sigma_U$ in (\ref{NormalApproxTDC}) by $\sigma_U(\hat{\eta})$, where $\hat{\eta}$ is the MLE of the copula parameter $\eta$.

\pagebreak
As the Pareto copula used by \cite{Schmidt2006} to approximate the lower tail dependence coefficient is upper-tail independent, we instead use the well-known Gumbel--Hougaard copula
\begin{equation} \label{GHCopula}
C_\eta(u,v) = \exp \left(- \Bigl[ \bigl(-\log(u)\bigr)^\eta + \bigl(-\log(v)\bigr)^\eta \Bigr]^{\frac{1}{\eta}} \right), \quad \eta \in [1,\infty),
\end{equation}
see, e.g., \cite{Nelsen2006}. For the rest of this section, we assume $\eta > 1$, as the Gumbel--Hougaard copula with $\eta = 1$ degenerates to the independence copula, which is not suited\footnote{One can easily check the intuition that $\Lambda_U(x,y) \equiv 0$ and thus $\sigma_U^2 = 0$ in this case.} for (\ref{NormalApproxTDC}). 

To derive $\sigma_U^2(\eta)$, we first note that for any Archimedean copula
\begin{equation*}
C_\eta(u,v) = \varphi_\eta^{-1}\bigl( \varphi_\eta(u) + \varphi_\eta(v) \bigr), \quad (u,v) \in (0,1)^2,
\end{equation*}
with suitable generator $\varphi_\eta$, we can use l'Hôpital's rule to write
\begin{align*}
\lim_{t\to \infty} t \left[ C_\eta\left(1 - \frac{x}{t},1 - \frac{y}{t} \right) - 1\right] &= \lim_{t\to \infty} \frac{ \varphi_\eta^{-1}\left(\varphi_\eta\bigl(1 - \frac{x}{t}\bigr) + \varphi_\eta\bigl(1 - \frac{y}{t}\bigr) \right) - 1}{t^{-1}} \\
&= - \lim_{t\to \infty} \frac{ x \varphi'_\eta\left(1 - \frac{x}{t}\right) + y \varphi'_\eta\left(1 - \frac{y}{t}\right)}{ \varphi_\eta'\left( \varphi_\eta^{-1}\Bigl( \varphi_\eta\bigl(1 - \frac{x}{t}\bigr) + \varphi_\eta\bigl(1 - \frac{y}{t}\bigr) \Bigr) \right)}\, .
\end{align*}
Setting $\varphi_\eta(u) = \bigl[-\log(u) \bigr]^\eta$ to match (\ref{GHCopula}), the tail copula is thus given by
\begin{align*}
\Lambda_{U,\eta}(x,y) &= x + y - \lim_{t\to \infty} \frac{ x \varphi'_\eta\left(1 - \frac{x}{t}\right) + y \varphi_\eta\left(1 - \frac{y}{t}\right)}{ \varphi_\eta'\left( \varphi_\eta^{-1}\Bigl( \varphi_\eta\bigl(1 - \frac{x}{t}\bigr) + \varphi_\eta\bigl(1 - \frac{y}{t}\bigr) \Bigr) \right)} \\
&= x + y - \lim_{t\to \infty} \left\{ \frac{x}{\left( 1 + \left[\frac{\log\left(1 - \frac{x}{t}\right)}{\log\left(1 - \frac{y}{t}\right)} \right] \right)^{1 - \frac{1}{\eta}}} + \frac{y}{\left( 1 + \left[\frac{\log\left(1 - \frac{y}{t}\right)}{\log\left(1 - \frac{x}{t}\right)} \right] \right)^{1 - \frac{1}{\eta}}} \right\} \\
&\underset{(*)}{=} x + y - \frac{x}{\Bigl[1 + \bigl(\frac{x}{y}\bigr)^\eta \, \Bigr]^{1 - \frac{1}{\eta}}} - \frac{y}{\Bigl[1 + \bigl(\frac{y}{x}\bigr)^\eta \, \Bigr]^{1 - \frac{1}{\eta}}} \\
&= x \left(1 - \Bigl[1 + \Bigl(\frac{x}{y}\Bigr)^\eta \, \Bigr]^{\frac{1}{\eta} - 1} \right) + y \left(1 - \Bigl[1 + \Bigl(\frac{y}{x}\Bigr)^\eta \, \Bigr]^{\frac{1}{\eta} - 1} \right),
\end{align*}
where $(*)$ follows by applying l'Hôpital's rule to derive
\begin{equation*}
\lim_{t\to \infty} \frac{\log\left(1 - \frac{y}{t}\right)}{\log\left(1 - \frac{x}{t}\right)} = \lim_{t\to \infty} \frac{\bigl(\frac{y}{t^2}\bigr)\left( 1 - \frac{x}{t} \right)}{\bigl(\frac{x}{t^2}\bigr)\left( 1 - \frac{y}{t} \right)} = \frac{y}{x} \, , \quad \lim_{t\to \infty} \frac{\log\left(1 - \frac{x}{t}\right)}{\log\left(1 - \frac{y}{t}\right)} = \frac{x}{y}\, . 
\end{equation*}
Thus, the partial derivatives\footnote{Note that $\Lambda_{U,\eta}(x,y)$ inherits the symmetry of $C_\eta(u,v)$.} are given as
\begin{align*}
\frac{\partial}{\partial x}\Lambda_{U,\eta}(x,y) &= 1 - \Bigl[ 1 + \Bigl(\frac{y}{x}\Bigr)^\eta \Bigr]^{\frac{1}{\eta} - 1} + x\left[ \frac{\eta - 1}{\eta} \Bigl(1 + \Bigl(\frac{y}{x}\Bigr)^\eta \Bigr)^{\frac{1}{\eta} - 2}(-\eta)\frac{y^\eta}{x^{\eta + 1}} \right]\\
&\qquad + y\frac{\eta - 1}{\eta}\Bigl(1 + \Bigl(\frac{x}{y}\Bigr)^\eta \Bigr)^{\frac{1}{\eta} - 2}\eta \frac{x^{\eta - 1}}{y^\eta} \\
&= 1 - \Bigl[ 1 + \Bigl(\frac{y}{x}\Bigr)^\eta \Bigr]^{\frac{1}{\eta} - 1} - (\eta - 1)\Bigl(\frac{y}{x}\Bigr)^\eta \Bigl[ 1 + \Bigl(\frac{y}{x}\Bigr)^\eta \Bigr]^{\frac{1}{\eta} - 2} + (\eta - 1)\Bigl(\frac{x}{y}\Bigr)^{\eta-1} \Bigl[ 1 + \Bigl(\frac{x}{y}\Bigr)^\eta \Bigr]^{\frac{1}{\eta} - 2}, \\
\frac{\partial}{\partial y}\Lambda_{U,\eta}(x,y) &= 1 - \Bigl[ 1 + \Bigl(\frac{x}{y}\Bigr)^\eta \Bigr]^{\frac{1}{\eta} - 1} - (\eta - 1)\Bigl(\frac{x}{y}\Bigr)^\eta \Bigl[ 1 + \Bigl(\frac{x}{y}\Bigr)^\eta \Bigr]^{\frac{1}{\eta} - 2} + (\eta - 1)\Bigl(\frac{y}{x}\Bigr)^{\eta-1} \Bigl[ 1 + \Bigl(\frac{y}{x}\Bigr)^\eta \Bigr]^{\frac{1}{\eta} - 2}.
\end{align*}
Using $\frac{\partial}{\partial x}\Lambda_{U,\eta}(1,1) = \frac{\partial}{\partial y}\Lambda_{U,\eta}(1,1) = 1 - 2^{\frac{1}{\eta} - 1}$ and rearranging yields 
\begin{equation*}
\sigma_U^2(\eta) = \left(2^{\frac{1}{\eta} + 1} - 2^{\frac{2}{\eta}} \right)\left(2^{\frac{1}{\eta} - 1} - \frac{1}{2} \right),
\end{equation*}
which finally results in the approximate 95\% confidence interval for $\lambda^{\circ,*}_U$ being given as
\begin{equation*}
\left[ \hat{\lambda}_U^{\circ,*} -z_{0.975}\sqrt{\frac{\left(2^{\frac{1}{\eta} + 1} - 2^{\frac{2}{\eta}} \right)\left(2^{\frac{1}{\eta} - 1} - \frac{1}{2} \right)}{k}}, \hat{\lambda}_U^{\circ,*} + z_{0.975}\sqrt{\frac{\left(2^{\frac{1}{\eta} + 1} - 2^{\frac{2}{\eta}} \right)\left(2^{\frac{1}{\eta} - 1} - \frac{1}{2} \right)}{k}} \; \right],
\end{equation*}
where $z_p$ is the $p$-quantile of the $\mathcal{N}(0,1)$ distribution.
\end{appendix}


\newpage
\bibliography{Optimal_BR_Penalty}

\end{document}